\documentclass[sn-mathphys]{sn-jnl}
\jyear{2023}%

\theoremstyle{thmstyleone}%

\theoremstyle{thmstyletwo}%

\theoremstyle{thmstylethree}%

\newcommand{\blue}[1]{\color{blue}#1}

\newcommand{\ie}{i.e., }
\newcommand{\eg}{e.g., }
\usepackage{booktabs}
\usepackage{dirtytalk}

\begin{document}

\title[Band-gap regression with architecture-optimized MPNNs]{Band-gap regression with architecture-optimized message-passing neural networks}

\author*[1,2]{\fnm{Tim} \sur{Bechtel}}\email{tim.bechtel@physik.hu-berlin.de}

\author[1,2]{\fnm{Daniel T.} \sur{Speckhard}}\email{daniel.speckhard@physik.hu-berlin.de}

\author[3,1]{\fnm{Jonathan} \sur{Godwin}}\email{jonathan@orbitalmaterials.com}

\author[1,2]{\fnm{Claudia} \sur{Draxl}}\email{claudia.draxl@physik.hu-berlin.de}

\affil*[1]{ \orgname{Humboldt-Universit\"at zu Berlin}, \orgaddress{\street{Zum Gro\ss en Windkanal 2}, \postcode{12489} \city{Berlin},  \country{Germany}}}

\affil[2]{\orgname{Max Planck Institute for Solid State Research}, \orgaddress{\street{Heisenbergstraße 1}, \postcode{70569} \city{Stuttgart},  \country{Germany}}}

\affil[3]{\orgname{Orbital Materials}, \city{London}, \country{United Kingdom}}

\abstract{Graph-based neural networks and, specifically, message-passing neural networks (MPNNs) have shown great potential in predicting physical properties of solids. In this work, we train an MPNN to first classify materials through density functional theory data from the AFLOW database as being metallic or semiconducting/insulating. We then perform a neural-architecture search to explore the model architecture and hyperparameter space of MPNNs to predict the band gaps of the materials identified as non-metals. The parameters in the search include the number of message-passing steps, latent size, and activation-function, among others. The top-performing models from the search are pooled into an ensemble that significantly outperforms existing models from the literature. Uncertainty quantification is evaluated with Monte-Carlo Dropout and ensembling, with the ensemble method proving superior. The domain of applicability of the ensemble model is analyzed with respect to the crystal systems, the inclusion of a Hubbard parameter in the density functional calculations, and the atomic species building up the materials.}

\keywords{graph neural networks, band gap prediction, neural architecture search, uncertainty quantification, density functional theory}



\maketitle

\section{Introduction}\label{sec1}
The success of density functional theory (DFT) has allowed researchers to predict material properties outside of the laboratory. There are several materials databases, such as NOMAD (Novel Materials Discovery)~\cite{draxl2019nomad}, the Materials Project \cite{jain2013commentary}, AFLOW (Automatic Flow of Materials Discovery)~\cite{curtarolo2012aflow}, OQMD (The Open Quantum Materials Database)~\cite{saal2013OQMD}, and others, collecting DFT data. NOMAD, for instance, contains over 140 million ground-state calculations. These databases have not only allowed researchers to avoid performing the same calculations again and again, thus saving computational resources, but have also enabled the re-purposing of data. For instance, one can train statistical-learning models, \eg neural networks (NNs) \cite{jha2018elemnet}, to predict DFT results with great accuracy~\cite{kulik2022roadmap}.

In order to infer properties of solid materials, details of the crystal structure are essential. In graph neural networks (GNNs), the geometrical information, \ie unit cell and atomic basis, can effectively be fed into ML models by representing the atoms as nodes and the atomic distances as the edges between the nodes. GNNs have seen success in predicting bulk properties of materials using Materials Project and OQMD data~\cite{xie2018crystal}. In this work, we start with a message-passing neural network (MPNN) as described in Ref. ~\cite{jorgensen2018neural}. This MPNN learns a representation for each atomic element in the first layer of the network, the embedding. The embedding is then iteratively updated for each atom using information from neighboring nodes. While the original implementation \cite{jorgensen2018neural} was in Tensorflow, we make use of the Jraph library~\cite{jraph2020github} developed at Deepmind, verifying our results with the QM9 \cite{ramakrishnan2014quantum} and Materials Project datasets. Our main efforts are, however, focused on the AFLOW database~\cite{curtarolo2012aflow} where both formation energies and band gaps are available for 62~\!102 structures. We train an MPNN to classify these materials as being metals or non-metals, and predict the DFT band gaps of the latter. The MPNN is also used to predict the formation energies of these materials. We perform a random-search-based neural architecture search on our network over various hyperparemeters, like number of MP steps, latent size, and learning rate, and demonstrate the complicated combined effect of these architecture parameters on the network performance. The best ten models by performance on the validation split are pooled in an NN ensemble~\cite{dietterich2000ensemble} to average their predictions on individual structures, which results in average predictions better than the best single model and existing models in the literature. Moreover, the domain of applicability of the model is analyzed with respect to the atomic composition, the crystal structure, and the inclusion of a Hubbard parameter in the calculation. Finally, we evaluate the possibility of using standard deviations from the model predictions, by comparing the ensemble with Monte-Carlo dropout (MCD), as a means of providing the user with uncertainty estimates on single predictions.

\section{Methods}\label{sec11}

\subsection{Graph representation of solids}
\begin{figure}[h]
\centering
\includegraphics[scale=0.22]{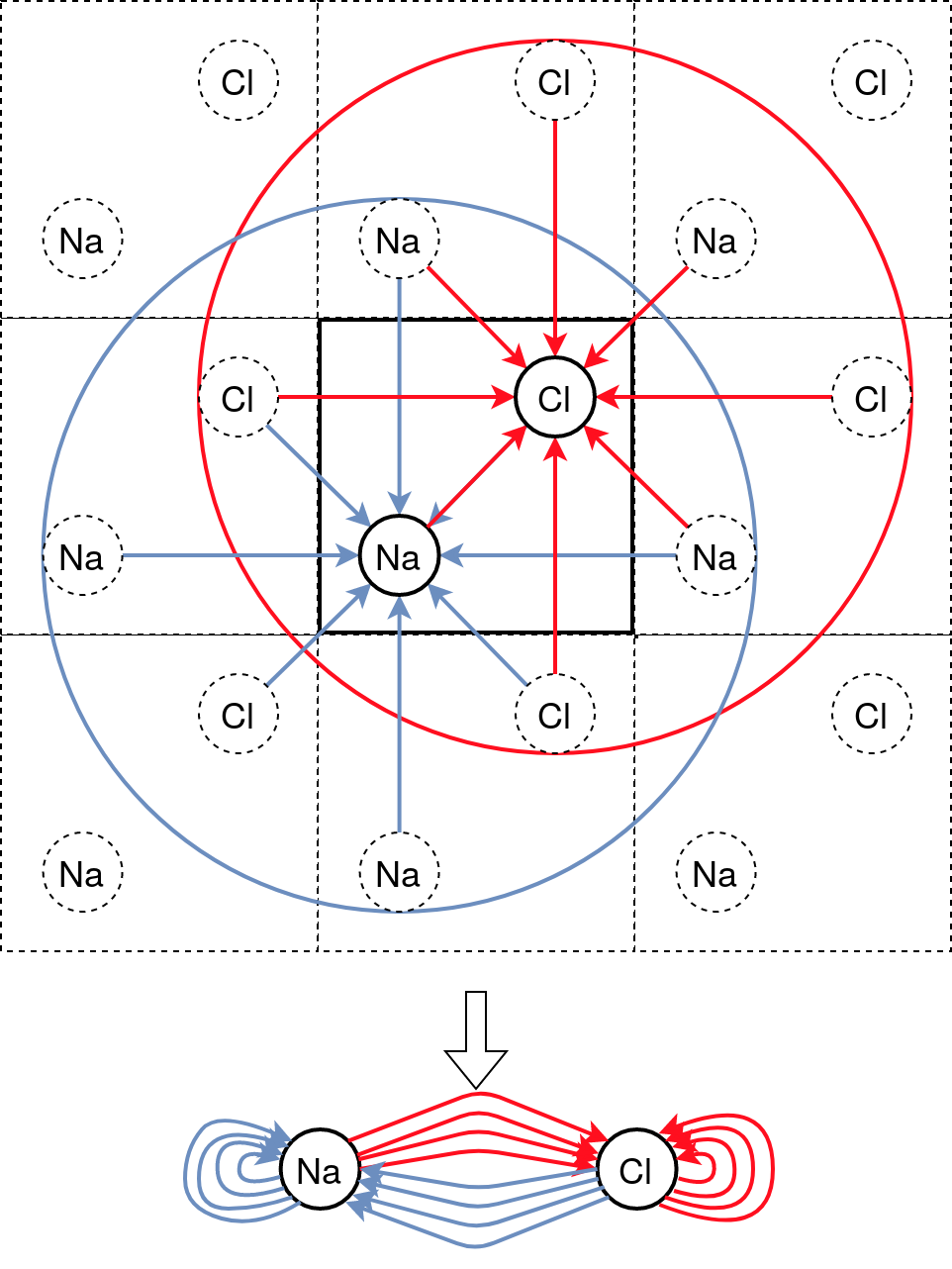}
\caption{Construction of a graph representation (bottom)  from the  two-dimensional periodic unit cell of NaCl (top) using a fixed cutoff radius. The unit cell in the center is indicated by the bold square. Neighboring unit cells are also shown (dotted squares). The cutoff radii for Na and Cl, are shown with a red and blue circle, respectively, centered on their atomic positions. }
\label{fig:unit cell graph}
\end{figure}

A crystalline material is described by a periodically repeated unit cell. It is fully characterized by the lattice vectors, the involved atoms, and their positions in the unit cell. Graphs allow one to construct representations of physical systems that have  translational and rotational invariance and are thus well suited for our purpose. In a graph representation, the local neighborhood of each atom can be defined by the distance to every other atom within a specified cutoff radius. When we apply this to a crystalline material, the cutoff radius may extend beyond its unit cell. An example of this can be seen in Fig. \ref{fig:unit cell graph} for the graph construction of two-dimensional NaCl. As all atoms within the unit cell, \ie one Na and one Cl atom, represent {\it nodes}, we end up with two nodes in our example (bottom panel). All atoms within the respective circle are then connected to this respective atom via {\it edges}. Here, the Na node is connected with edges to four Cl atoms and four Na atoms in neighboring unit cells. The latter links are called {\it self-edges} that encapsulate the periodic boundary conditions of the crystal in the resulting graph, even though the graph representation itself is not explicitly periodic. Note that this graph construction uses directional edges where the edge originates from a node and ends on a node. Directional edges are used throughout this paper. Their use enables asymmetric graph representations, like a k-nearest neighbor graph, which is often favorable.

Once the nodes and edges of the graph are determined, we can define the adjacency matrix as: 
\begin{equation}
  A_{ij} =
    \begin{cases}
      1 & \text{if node $i$ is connected to node $j$}\\
      0 & \text{if $i=j$ or otherwise}
    \end{cases}      
    \label{eq:adjacency matrix}
\end{equation}
The neighborhood of node $i$ is then formally defined as
\begin{equation}
    N(i) = \{ j \mid A_{ij} = 1\}.
   \label{eq:neighborhood}
\end{equation}
It is not possible to construct the adjacency matrix for periodic systems, as there can be multiple edges for a pair of atoms. Such a representation is not considered a simple graph (which has at most one edge per pair of nodes); rather, it is considered a multi-graph (see Fig. \ref{fig:unit cell graph}). However, we can still define the neighbourhood of a node $i$ by the set of other nodes to which it is connected.
Apart from using a constant cutoff to define the atomic neighborhood, the edges can also be constructed by considering a fixed number ($k$) of nearest neighbors (termed KNN algorithm). When the cutoff radius is the same for all atoms in the unit cell, the resulting graph is symmetrical. However, in the KNN algorithm, the constructed graph is not necessarily symmetrical but each node in the graph has the same number of neighbors. The KNN approach has several benefits. The cutoff radius can, in principle, produce isolated nodes, \ie nodes without neighbors.
As a reference to our studies, we refer to an architecture ~\cite{jorgensen2018neural} that was optimized to fit formation energies of structures present in the Materials Project database~\cite{jain2013commentary} and OQMD~\cite{saal2013OQMD}, with the calculation of both being performed with the same DFT code and functional~\cite{hegde2020quantifying}. In that work, it has been shown that a KNN cutoff with $k=24$ neighbors produces the lowest mean absolute error (MAE) for an MPNN with edge updates when predicting formation energies on a OQMD materials dataset, but the improvement over $k=12$ neighbors is only marginal. In our experiments, training is sped up by about 20\% using the lower number or neighbors, while not affecting the model performance significantly. Thus, in our search for an optimal message-passing architecture, we adopt $k=12$, since --as we will show further below-- this helps to reduce our already very large search space.

\subsection{Message-passing algorithm}
MPNNs and, more generally, GNNs work by iteratively updating hidden graph states. They contain information concerning the atoms in the material and their interactions. In this work, we use hidden node and edge states which we refer to as {\it node and edge vectors}, respectively. Each hidden vector has its own update equation, as described in Section \ref{sec: update equations}. They are the same as used in Ref.~\cite{jorgensen2018neural}. After a fixed number of updates ($L$), the node vectors are fed into a readout function that predicts the target property.

\subsubsection{Node and edge embeddings} \label{sec: embedding}

The raw node and edge features are first transformed into representations that facilitate the graph network to learn from the input data. The atomic number, $Z_i$, of each node $i$ is one-hot encoded (OHE) into a vector. Its dimension is the number of different atomic species in the dataset. For instance, in a dataset containing 74 different elements of the Periodic Table of Elements (PTE), each atom is assigned a 74-dimensional binary vector, where only a single bit is non-zero (\eg hydrogen is represented with $ [1, 0,0,\dots,0]$ and helium with $ [0, 1,0,\dots,0]$). The element type, represented as an OHE vector is then transformed into a vector with latent size $C$ by multiplication with a trainable weight matrix $W_0$:
\begin{equation}
    h_i^0=W_0 \cdot \text{OHE}(Z_i) .
\end{equation}
With $W_i$, we denote the weight matrices, whose elements are optimized during the training process.

Each edge in our graph is represented by a hidden edge-vector state. The initial state (embedding) is computed by feeding the pairwise atomic distance, $d_{ij}$, between two nodes, $i$ and $j$, into a basis-function expansion. As $d_{ij}$ is always translationally and rotationally invariant, the graph reflects this desired property. In this work, we use Gaussian basis functions:
\begin{equation}
(e_{ij})_{\kappa} = exp\left\{-\frac{\left[d_{ij} - (- \mu_{min} - \kappa\Delta)\right]^2}{\Delta}\right\}, \quad \kappa=0\dots \kappa_{max}-1.
\label{eq:readpit function}
\end{equation}
The parameters $\mu_{min}$ (offset of the basis functions), $\Delta$ (width of the basis functions), and $\kappa$ are chosen to span the range of input features. In Ref. \cite{jorgensen2018neural}, $\mu_{min}$ is set to 0 \AA, $\Delta$ to 0.1 \AA, and $\kappa_{max}$ to 150. We use these values in this work as well. This dimensional expansion of the scalar distance into a vector of size $\kappa_{max}$ might seem strange but is analogous to OHE in that the model is then able to decorrelate input and output more easily with the transformed, now higher-dimensional input~\cite{jorgensen2018neural}.

\subsubsection{Node/edge update functions}
\label{sec: update equations}
The nodes and edges are updated at each message passing (MP) step $l$. First, the edge update is performed, then the node update is applied using the updated edges. For each edge, $M_{ij}$ is defined as an edge-wise message that connects node $i$ to node $j$ (node $j$ is sending, and node $i$ is receiving the {\it message}):
\begin{equation}
    M_l(h_i^l, h_j^l, e_{ij}^l)=M_l(h_j^l,e_{ij}^l)=(W_1^l h_j^l)\odot \sigma (W_3^l\sigma(W_2^l e_{ij}^l)).
\label{eq:message edge function}
\end{equation}
Here, the symbol $\odot$ denotes element-wise multiplication, and $\sigma$ an arbitrary, non-linear activation function. The element-wise multiplication can be seen as a continuous filter, where the edge feature attenuates the node feature, after both have been transformed by feed-forward layers.

Edge-wise messages are aggregated into node-wise messages by either taking the sum of neighboring features as in
\begin{equation}
    m_i^{l+1}=\sum_{j\in N(i)} M_l(h_i^l, h_j^l, e_{ij}^l)
\label{eq:message aggregation}
\end{equation}
or any permutation-invariant aggregation function (\eg mean, minimum, maximum, etc.)~\cite{corso2020principal}. The edge-update function consists of concatenating the sending and receiving nodes with the edge feature $e_{ij}$. This concatenation is then passed into a two-layer NN with two shifted soft-plus activation functions,
\begin{equation}
    e_{ij}^{l+1}=\sigma (W_{l+1}^{E2}\sigma(W_{l+1}^{E1}(h_i^{l+1};h_j^{l+1};e_{ij}^l))) .
    \label{eq:edge update function}
\end{equation}
Nodes are then updated according to 
\begin{equation}
    h_i^{l+1}=S_t(h_i^l,m_i^{l+1})=h_i^l+W_5^l\sigma (W_4^l m_i^{l+1})
\label{eq:state transition function}
\end{equation}
by using the aggregated messages $m_i^{l+1}$ and the original node features $h_i$. The node-wise message is transformed in a two-layer NN with an activation function and is added to the previous node feature $h_i^{l}$, to arrive at the updated node feature $h_i^{l+1}$. This addition has similarities to the residual connections used in ResNet architectures, which enable training of deeper NNs~\cite{he2016resnet}.

\subsubsection{Global readout function} \label{sec: readout function}
After $L$ MP steps, the procedure is stopped, and the node features are aggregated into a single scalar, transforming them by means of an NN with two layers and a hidden size of $C/2$. Subsequently, one sums over all nodes in the graph or takes the mean. This step is required since the aggregation should be invariant with respect to the permutation of nodes, as their ordering should not matter. Whether the sum or the average is taken over all nodes depends on the dataset and the target property. Here, we show the equation for the summation:
\begin{equation}
    R(\{h_i^L\in G\})=\sum_{h_i^L\in G} W_7 \ \! \sigma (W_6 h_i^L)
\label{eq:readout}
\end{equation}
As the graphs can have variable sizes, the aggregation should also be able to handle varying numbers of nodes in the graph. Note, for the QM9 dataset, where the target is the total internal energy $U_0$, we use a sum in the readout function. For the datasets, where the formation energy per atom is targeted, we take the mean. For further discussion on readout aggregation methods, see Ref.~\cite{gong2022examining}. 

Combining the pieces of the node/edge embedding, the node/edge update functions, and the readout function, we arrive at the complete algorithm for a message-passing edge-update neural network, abbreviated as MPEU:

\begin{algorithm}
\caption{Message-passing algorithm with edge updates.}\label{alg:MPEU}
\begin{algorithmic}

\Function{GNN}{V,E}
\Comment{{\blue First, embed edges and nodes}}
\ForAll{$d_{ij}\in E$}
    \State $\varepsilon_{ij} \gets RBF(d_{ij})$ \Comment{{\blue Expand distances in radial basis functions}}
\EndFor

\ForAll{$Z_i \in V$} \Comment{{\blue Loop over atomic number of nodes in graph}}
    \State $h_i^1 \gets W_0 \cdot \text{OHE}(Z_i)$ \Comment{{\blue One hot encode elements and transform}}
\EndFor
\State $e_{ij}^{0}\gets E_0(h_i^{0}, h_j^{0}, \varepsilon_{ij})$
\Comment{{\blue Apply zeroth edge update layer}}
\State $l \gets 0$ \Comment{{\blue Initialize layer counter}}
\While{$l \le L$}
    \State $M_{ij} \gets M_l(h_i^{l},h_j^{l},e_{ij}^{l})$ \Comment{{\blue Calculate edge-wise messages}}
     \State $m_j^{l+1}\gets \sum_{i\in N(j)}M_{ij}$
     \Comment{{\blue Aggregate edge-wise messages to nodes}}
     \State $h_i^{l+1} \gets S_t(h_i^l,m_i^{l+1})$ \Comment{{\blue Update nodes with incoming messages}}
     \State $e_{ij}^{l+1}\gets E_l(h_i^{l+1}, h_j^{l+1}, e_{ij}^{l})$ \Comment{{\blue Update edges}}
     \State $l \gets l + 1$
\EndWhile
\State $\hat{y} \gets \frac{1}{N_N}\sum_{h_i^L \in G}\text{NN}(h_i^L)$ \Comment{{\blue Apply NN readout function}}
\State \Return{$\hat{y}$}
\EndFunction

\end{algorithmic}
\end{algorithm}

\subsection{Architecture search} \label{sec: method architecture search}
MPNNs as described above contain many architecture parameters. Not much work has been devoted to explore how they affect the model performance. In this work, we perform a neural architecture search (NAS) using a random search algorithm. We build our search space based on the MPNN model described in detail above, where the embedding and latent size (\eg the node/edge vector dimension), the number of MP steps, the activation function, the number of layers in the MLP in the node/edge update functions, and the number of layers in the readout NN are varied. Other MPNN parameters, such as the initial learning rate, the learning-rate decay, the batch size, the dropout and the layer norm are also varied concurrently, assuming that the importance of these variables is related to the number of parameters in the model. We note that the number of trainable weights, \ie parameters optimized during training, scales linearly with the number of MP steps, and quadratically with the latent size. The number of trainable parameters (weights) ranges from 500,000 up to 20,000,000, with the best models usually having around 1,000,000 weights. Neural architecture searches are often performed with a mix of explorative and exploitative algorithms such as Bayesian optimization or genetic algorithms~\cite{speckhard2023neural}. In this work, we opt to use a random search algorithm since we want to sample the large multidimensional space exploratively to gain a better understanding of the space. While Bayesian optimization and genetic algorithms sample the parameter space with a bias towards regions with well-performing models, random search samples the parameter space without bias, \ie purely exploratively.

\subsection{Neural-network ensembles} \label{sec:ensembles}
Given that a large number of models is trained in the process of the NAS, it is a natural step to not only look at the best performing model, but also at the predictions of the other models. If a number of well-performing and diverse models make a prediction on a single input, it can be expected that the average prediction of the models in the ensemble outperform the individual models~\cite{hansen1990neuralensembles,zhu2023uncertainty}. There are three main reasons for this~\cite{dietterich2000ensemble}: (i) Different models can achieve the same performance on the regression or classification task. An ensemble reduces the risk of choosing the poorly performing model when it is applied on the held-out test data. (ii) Due to the non-convex nature of optimizing a neural network, it is expected that training results in a local minimum with respect to the trainable parameters rather than the global minimum. This leads to the possibility of different locally optimal parameters given the same training data, if the initialization of the models is different. Again, the model ensemble reduces the risk of choosing a poorly performing model that is stuck in a local minimum far from the global minimum. (iii) By averaging across different models, the space of possible solutions is expanded, leading to an increased learning capacity of the model.

The main prerequisite for an ensemble to improve prediction quality is that the models are diverse, \ie are trained with different data and/or have different parameter values and or architectures, and that the different models are well performing on their own. For our ensemble, we select the ten best candidate architectures with respect to the validation dataset from our NAS. We expect that the variety of architectures and hyperparameter values for the different candidate models should result in the ensemble being less prone to over-fitting and performing better than the top NAS models individually.

\subsection{Uncertainty estimates} 
\label{sec: method MCD}

Reliable uncertainty estimates are important when deploying a machine-learning model in a real application \cite{busk2021calibrated}. They provide information on the model's domain of applicability so that the user can understand whether to trust an inference~\cite{sutton2020identifying}. Ensembling the top ten models from our NAS gives us a method to obtain an uncertainty estimate by looking at the predictions from all ten models in the ensemble and calculating the standard deviation. We compare this method with another popular method, Monte-Carlo dropout (MCD)~\cite{gal2016dropout}. Dropout means that nodes in the network are turned off/on probabilistically. In the case of MCD, the dropout is also used for model inferences (\ie nodes are turned off randomly for each prediction the model makes) and not just for training the model. This enables stochastic predictions from a {\it virtual ensemble}. Ideally, aggregating these predictions gives an uncertainty estimate in the same way as a Gaussian process would. We employ dropout on all NN layers in the model (readout function, edge-update function, etc.). Dropout also helps to prevent overfitting during training for regularization purposes~\cite{srivastava2014dropout}. The dropout is kept on for inferences; ten predictions are made for each input, of which the mean and standard deviation are reported.

\subsection{Band-gap classification and regression} \label{sec: method band gap classifier}

The Kohn-Sham band gaps obtained in density functional theory typically severely underestimate the corresponding quasi-particle gaps of the respective structures. This systematic error can be partly remedied by the use of hybrid functionals~\cite{xiao2011accurate}, but an expansive database has yet to be created using this method. It is expected, however, that a model that is fitted on biased data, carries the same bias during inference on unseen data. This should be kept in mind when discussing the use of GNNs trained on DFT data in high-throughput searches, \eg for large-band-gap materials. To make sure that we use as consistent a data set as possible, the data used in this work were filtered to only contain DFT calculations performed with the PBE~\cite{perdew1996generalized} functional. For more details on the data, see below.

\begin{figure}[h]
\centering
\includegraphics[scale=0.35]{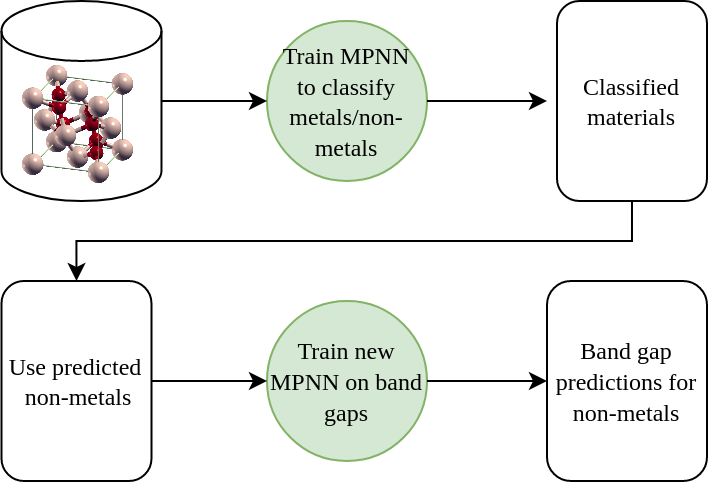}
\caption{Workflow of band-gap classification in terms of metals/non-metals using an MPNN, followed by the prediction of band gaps using another MPNN.}
\label{fig:egap classify flow}
\end{figure}

Following the literature on predicting band gaps on AFLOW data ~\cite{isayev2017universal}, we train two separate models. The first one classifies materials as non-metals and metals (having a zero DFT band gap), using a binary cross-entropy loss. The second model is fitted to predict the band gaps of the materials classified as non-metals. This workflow is illustrated in Fig. \ref{fig:egap classify flow}. Both models --classification MPNN and regression MPNN-- have a similar architecture. Since we find very high accuracy on the classification task without tuning the hyperparameters, we only perform a NAS on the band-gap regression task.

\subsection{Dataset} \label{sec: aflow_data}

For band-gap prediction and classification, we use all materials in AFLOW that have a band gap. The AFLOW data are obtained with the DFT code VASP~\cite{kresse1996VASP}. As mentioned above, we ony use those calculated with the PBE functional, and duplicates have been removed from the dataset. To simplify our analysis, we use the same dataset for the prediction of formation energies. That means that there are some materials for which a formation energy has been calculated but no band structure, and they are therefore excluded from the formation-energy regression.

Outliers with a formation energy of less than -10 eV/atom (\ie two materials, S and SiO$_2$, both space group 70) or higher than 70 eV/atom (a single material, BrIPb, space group 59) were removed. These outliers have formation energies more than five $\sigma$ away from the mean in the dataset. In this dataset, we have 46,090 metals and 16,012 non-metals. The dataset is therefore biased towards metals.

\subsection{PLMF and ElemNet Models}

To evaluate and compare the performance of the models from our NAS, we include several models from the literature. In the PLMF model~\cite{isayev2017universal}, the lattice structure was decomposed into fragments, and a ML model was trained on these fragments. To the best of our knowledge, the PLMF model is the only model in the literature that has been used to classify band structures and regress band gaps for the AFLOW dataset. One caveat to the comparison with this work is that neither the training/test splits, nor the code for the method employed in their work were shared in their original paper. We compare our results to the metrics reported in Ref.~\cite{isayev2017universal}, however, we note that their dataset is from an earlier snapshot of AFLOW. Since the publishing of that article, the AFLOW database has grown and some data-points have been recomputed (\eg with a newer version of the VASP code). More specific, we used the online API~\cite{gossett2018aflowml} to get results from their model trained with the earlier AFLOW dataset snapshot, evaluated on the current AFLOW test dataset. Therefore, not having access to the training/test splits used in Ref. ~\cite{isayev2017universal}, some of the test data might have already been seen by the trained PLMF.

We also compare our results with the deep neural network ElemNet~\cite{jha2018elemnet}, a model that does not use any structural information, but is given only the chemical formula (stoichiometry). This model was trained and evaluated using OQMD~\cite{saal2013materials} data and demonstrated good performance for formation energies. We retrain the model on AFLOW data but keep the model architecture from the original publication.

\section{Software implementation}
\label{section:software_implementation}

For our computational framework, the JAX ecosystem is used because of its frequent use in recent state-of-the-art research~\cite{jax2018github, godwin2021very}. Features like automatic differentiation and just-in-time compilation, and active development that foster new scientific discoveries are especially appealing. In conjunction with JAX, we use the Haiku library for trainable NN layers and Optax for optimization routines which is itself built upon JAX~\cite{haiku2020github, optax2020github}. Finally, to encode our MPNN architecture and update equations we use Jraph which provides a functional API to apply transformations to arbitrary graphs~\cite{jraph2020github}. These four libraries form a cohesive framework for the whole process of training a graph-based machine-learning model, apart from the database interface that we implemented.

As a local database for atomic structures, we employ the Atomic-Simulation-Environment database (ASE-DB)~\cite{larsen2017atomic}. The conversion of atomic structures to graphs is done only once for each database using a maximum neighborhood of k-nearest neighbors, therefore the time-consuming graph generation does not have to be repeated for each hyper-parameter experiment.

The data are divided into training, validation, and test data in an 80:10:10 split. Training and validation data are used for cross-validation and early stopping, and the model is finally evaluated on the unseen test data to asses its performance on samples it has not yet encountered. Early stopping is implemented as described in \cite{jorgensen2018neural}, by checking if the validation loss has decreased compared to the loss 1 million steps before. The model is trained with dynamic batches of a maximum of $N_{batch}$ graphs (including a padding graph) \cite{Speckhard2023batches} using the Adam optimizer provided by Optax~\cite{kingma2014adam}. Batches are sampled without replacement from the training dataset and reshuffled in every epoch. Dynamic batches are created by calculating the average number of nodes and edges for $N_{batch}-1$ graphs and rounding this result up, in this case to the next multiple of 64. This value (power of 2) is motivated by the processor architecture that is used, in that GPUs use banked memory and specific optimized kernels that work best with data sizes of $2^N$. Then, during the training loop, graphs are sampled without replacement from the training dataset, until the maximum number of nodes or edges is reached, or $N_{batch}-1$ graphs are retrieved. The rest of the budget is then used for padding, and the result is a static number of nodes, edges, and graphs. This only needs to be just-in-time compiled once and therefore greatly increases the speed of each graph network evaluation.

The implementation is validated by training a model on formation energies from the Materials Project, specifically the MP-crystals-2018.6.1 snapshot provided in~\cite{chen2019megnet}. With this training data, we obtain similar error metrics to Ref. \cite{jorgensen2018neural}, despite using a different training/test split (no information on the split used was provided in Ref. \cite{jorgensen2018neural}).

\section{Results}\label{sec: results}

We first train a model to classify materials as metals or non-metals, minimizing the binary cross-entropy. We evaluate the receiver operating characteristic (ROC) alongside the total accuracy of the model, since these two metrics are easier to interpret than the binary cross-entropy. The area under the curve (AUROC) of the receiver operator characteristic presents a balanced scalar metric of the classifier's performance. Recall, that an AUROC of one indicates a perfect fit, whereas an AUROC of 0.5 points at random predictions. The reference MPEU model performs quite well for the classification task with an accuracy of 0.98 and an AUROC over 0.99 as shown in Table \ref{table:1}. This is remarkable since the reference MPEU model has, to the best of our knowledge, not been trained to classify band gaps. Both the AUROC and accuracy are higher than the corresponding values for the PLMF model from the literature that are included for comparison. The high AUROC value indicates a very balanced classification performance across metals and non-metals, despite the dataset being biased towards metals. As a result of the satisfactory performance of the MPEU reference model, we decided not to perform further optimizations on the model with a NAS.

\begin{table}[h]
\centering
\caption{Summary of cross-validated models on formation energies ($E_f$) and band gaps ($E_g$) by our best performing MPEU model from the NAS. For comparison, literature results for the PLMF model on an earlier AFLOW dataset snapshot (marked by the $*$) are shown. Results of the classification into metals/non-metals are shown at the bottom.}
\fontfamily{cmr}\selectfont
\begin{tabular}{p{2.7cm} p{3.5cm} p{1.1cm} p{1.0cm} p{1.0cm}}
\toprule
 \textbf{Property} & \textbf{Model} & \textbf{RMSE} & 
 \textbf{MAE} & \textbf{MdAE} \\
 \midrule
 $E_g$ $[$meV$]$  & Ensemble & \textbf{379} & \textbf{168} & \textbf{26.2} \\
 & Best in NAS & 469 & 205 & 35.0 \\
 & Reference~\cite{jorgensen2018neural} & 399 & 180 & 32.9  \\
 & SchNet~\cite{schutt2018schnet} & 489 & 235 & 68.4 \\
 & PLMF (new data)~\footnotemark[1] & 1327 & 618 & 151 \\
 & PLMF (reported)~\cite{isayev2017universal} & 510 & 350 & - \\
 & ElemNet~\cite{jha2018elemnet} & 816 & 515 & 303 \\
 \midrule
 $E_f$ $[$meV/atom$]$ & Ensemble & \textbf{56.3} & \textbf{15.0} & \textbf{6.29}\\
 & Best in NAS & 65.4 & 21.0 & 10.7 \\
 & Reference~\cite{jorgensen2018neural} & 57.5 & 17.9 & 8.32 \\
 & SchNet~\cite{schutt2018schnet} & 68.0 & 29.3 & 17.2 \\
 & ElemNet~\cite{jha2018elemnet} & 214 & 135 & 68.6 \\
 \midrule
 & & \textbf{Accuracy} & & \textbf{AUROC}\\
 \midrule
 Classification & Reference & \textbf{0.98} & & \textbf{$>$0.99} \\
  & PLMF (new data)~\footnotemark[1] & 0.97 & & - \\
  & PLMF (reported)~\cite{isayev2017universal} & 0.93 & & 0.98\\
 \bottomrule
\end{tabular}
\label{table:1}
\end{table}
\footnotetext[1]{The AFLOW dataset has grown since the evaluation was done in Ref. \cite{isayev2017universal}.}

\begin{figure}[h]
\centering
\includegraphics[scale=0.6]{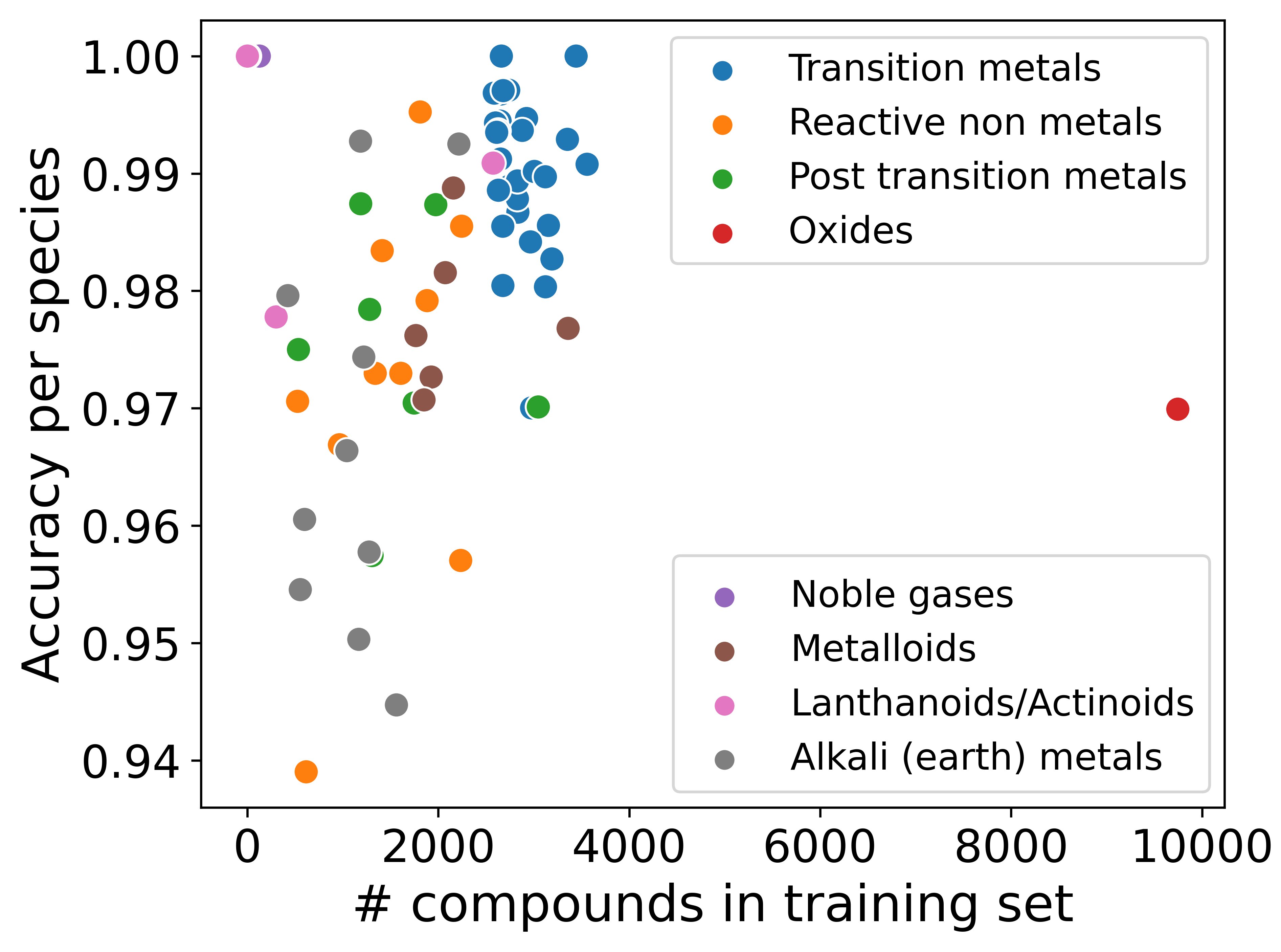}
\caption{Accuracy of the classification between metals/non-metals for different material classes evaluated on the test split.}
\label{fig:error_vs_species_class}
\end{figure}

In Fig. \ref{fig:error_vs_species_class}, the performance of the classifier in terms of accuracy is analyzed depending on how often each type of material appears in the training split. As expected, we see a general trend that the more often such category appears in the dataset, the higher the classification accuracy of the model is. For instance, transition metals appear frequently in the training split and are generally classified correctly with more than 98\% of the time. In contrast, the fewer alkali metals are classified with accuracies ranging from 94\% to 99\%. Oxides, however, are outliers in this trend. Despite being best represented in the training dataset with over ten thousand materials, they are classified with an accuracy of 97\%, lower than the mean accuracy over the entire dataset (98\%). One reason for this could be the fact that for many of the transition metal-oxides, a Hubbard-U correction has been applied in the production of the DFT data. We will come back to this point further below.

\begin{figure}[h]
\centering
\includegraphics[scale=0.30]{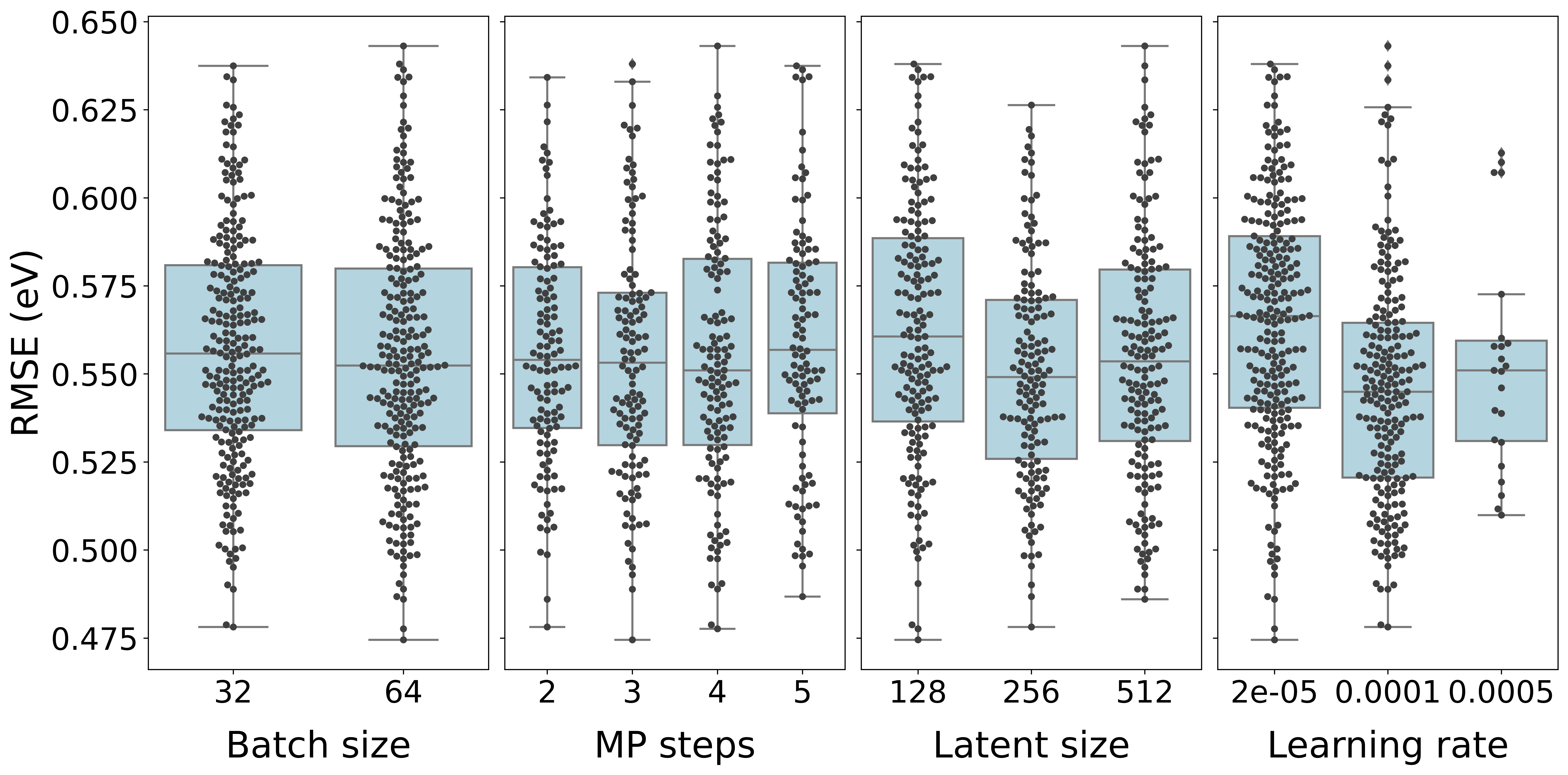}
\caption{Neural architecture search results for the band gap, using the non-metals as predicted by the classifier (trained on materials from the AFLOW database). The RMSE on the validation split are shown for model-architecture parameters and settings in our search space. The effect of several additional parameters is shown in the appendix (Fig. \ref{fig:grid_search_aflow_1_bandgap}).}
\label{fig:grid search aflow bandgap}
\end{figure}

After classification, we predict the band gaps of those materials that have been classified as non-metals. We visualize in Fig.~\ref{fig:grid search aflow bandgap} the results of the neural-architecture search for band-gap regression on the validation split. Additional hyperparameters are shown in the appendix (Fig.~\ref{fig:grid_search_aflow_1_bandgap}). The results indicate that, in general, the larger batch size of 64 is only slightly better than 32. Three message-passing steps give the lowest mean RMSE. A latent size of 256 is favored and a learning rate of 1E-4 is significantly preferred. Although, for instance, a latent size of 256 and a learning rate of 1E-5 are preferred on average, the best NAS model has a latent size of 128 and a learning rate of 2E-5 which shows the correlation between variables. We see that increasing the latent size and using a smaller learning rate, which should increase the model's learning capacity, does not result in a lower RMSE. This may hint at a rather small amount of training data for the problem. In total, 500 random models were trained; for 459, the loss converged and training was stopped early; for 41, the optimization was aborted due to an unstable loss value (as often observed by us when a higher learning rate is paired with no layer normalization). The NAS model that performed best on the validation dataset in terms of RMSE was selected as the model used for testing.

The regression metrics, RMSE, MAE, and median absolute error (MdAE), are collected in Table \ref{table:1}. The MdAE is an error metric that is unaffected by outliers, as it gives a midpoint where the same number of absolute errors lie above and below. The low MdAE value across models shows that both the RMSE and MAE are affected by outliers with a high absolute error. We observe that the best NAS architecture is overall similar to the reference model, however, the optimal learning rate at 2E-5 is lower, and the NAS model uses a dropout rate of 0.05 while the reference does not use any. The best NAS model actually performs worse than the reference model on which the NAS space is created. A table comparing the validation results and the test results is shown in Table~\ref{table:validation} in the appendix. 

The validation and test RMSE are very similar for the best NAS Model (468 eV and 469 eV, respectively). We conclude that the best NAS model is not overfitting the validation dataset despite our NAS choosing the best model based on the validation results. In contrast, for the reference model, the validation RMSE is much higher than the test RMSE (505 eV and 399 eV, respectively). This suggests that the reference model's superior test performance may depend on the split. Our ensemble model that combines the top ten NAS models, outperforms the reference model significantly in terms of MAE, RMSE, and MdAE. Note, that in order to evaluate all band gap regression models on equal footing, we use the same MPEU band-gap classifier so that each model uses the same training, validation, and test dataset.

\begin{figure}[h]
\centering
\includegraphics[scale=0.5]{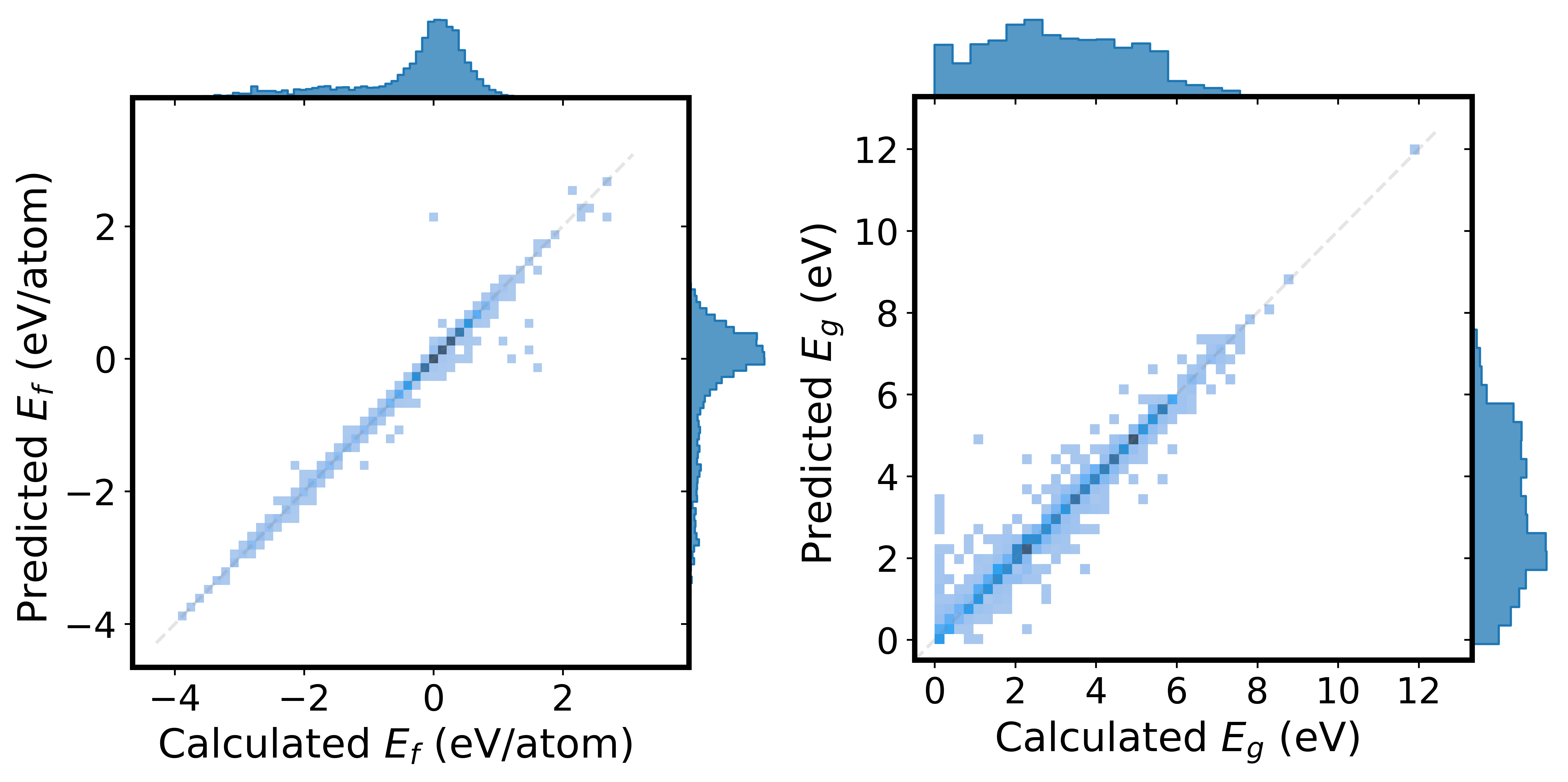}
\caption{Regression results using the ensemble MPEU model on AFLOW data for formation energies (left) and band gaps of predicted non-metals (right). Marginal histograms showing the distribution of predicted and calculated values are shown above the horizontal axes and to the right of the vertical axes.}
\label{fig:regression combined}
\end{figure}

\begin{figure}[h]
\centering
\includegraphics[scale=0.6]{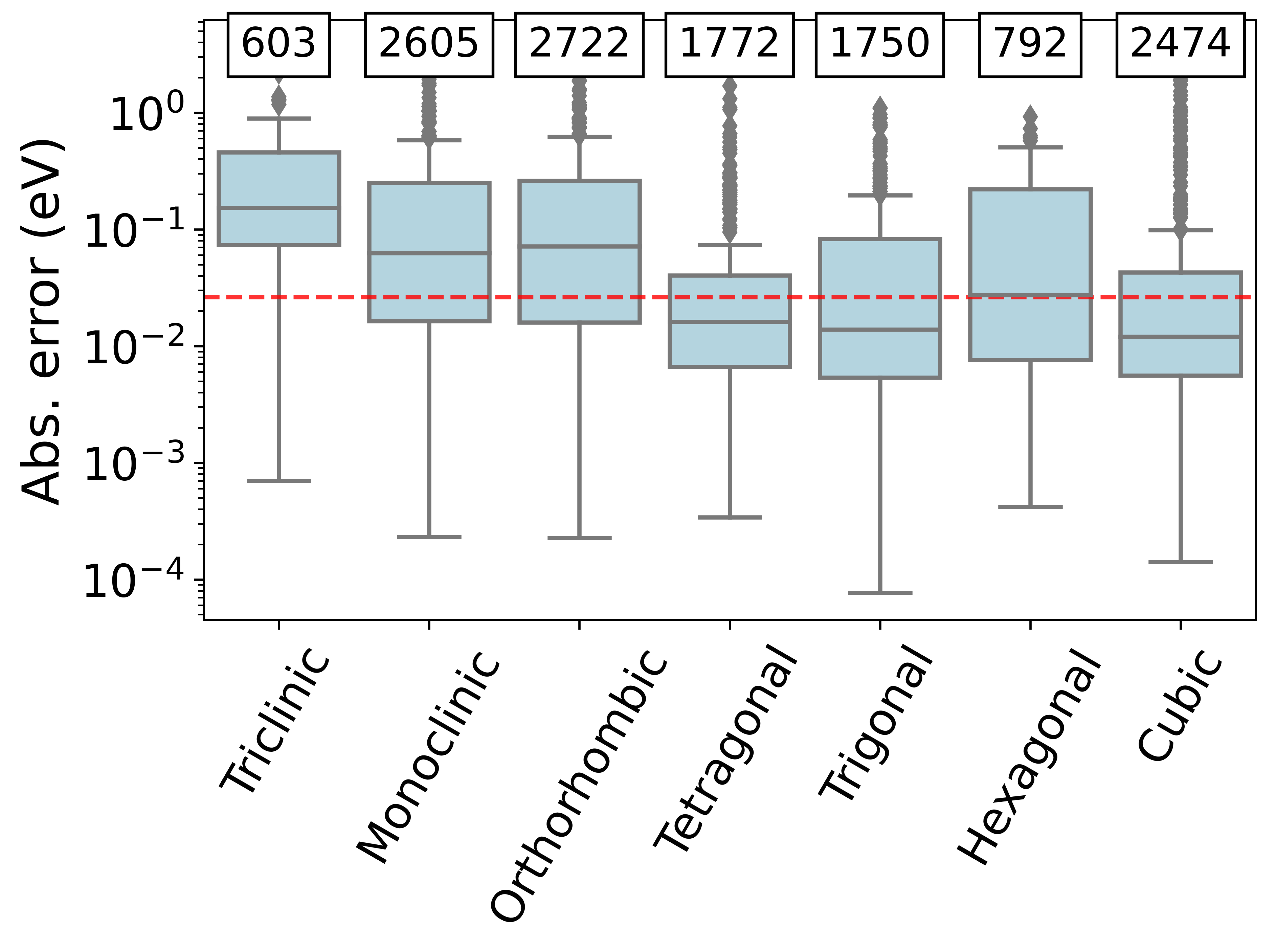}
\caption{Distribution of absolute errors in band gaps for different crystal systems using the NAS ensemble model. The number of training materials for each crystal system is displayed at the top. The dashed red line shows the MdAE. Points above the 95\% quantile are shown to better understand the distribution.}
\label{fig:error_vs_crystal_bandgap}
\end{figure}

Analyzing the performance of the ensemble model further, we see the largest RMSE occurs on data that our classification model incorrectly predicted as non-metals. This is seen in Fig. \ref{fig:regression combined}. All MPEU models significantly outperform the PLMF and ElemNet model indicating the superiority of graph-based models for this task and dataset. The crystal structure also plays an important role in how well the MPEU model for predicting band gaps performs, as seen in Fig. \ref{fig:error_vs_crystal_bandgap}. In general, we see a better band-gap prediction for lattice types with more materials in the training set. For instance, cubic systems are quite common and have the lowest median error while triclinic structures are the most rare and have a poor median error. That said, there is a similar number of hexagonal training structures compared to triclinic structures (792 vs 603) but we observe a much lower median band-gap error for the former. The dependence of the model performance on the lattice type indicates that the model is learning from the input crystal structure, which is desired.

We also train our models to predict formation energies, performing a NAS on this task. The results are shown in Table \ref{table:1}; the effect of several NAS parameters is shown in the appendix. They exhibit similarities to those of the band-gap regression. The ensemble NAS models performs the best, but the reference model out-performs the individual top-ranked NAS model. As the reference model~\cite{jorgensen2018neural} was trained on  formation energies of the Materials Project, it is no surprise that its architecture transfers very well to another dataset based on the same code. The two databases show, however, significant differences in computational details, such as convergence criteria for geometry optimization, or the use of DFT+U, and alike, as analyzed in detail in Ref. \cite{hegde2020quantifying}).

\begin{figure}[h]
\centering
\includegraphics[scale=0.6]{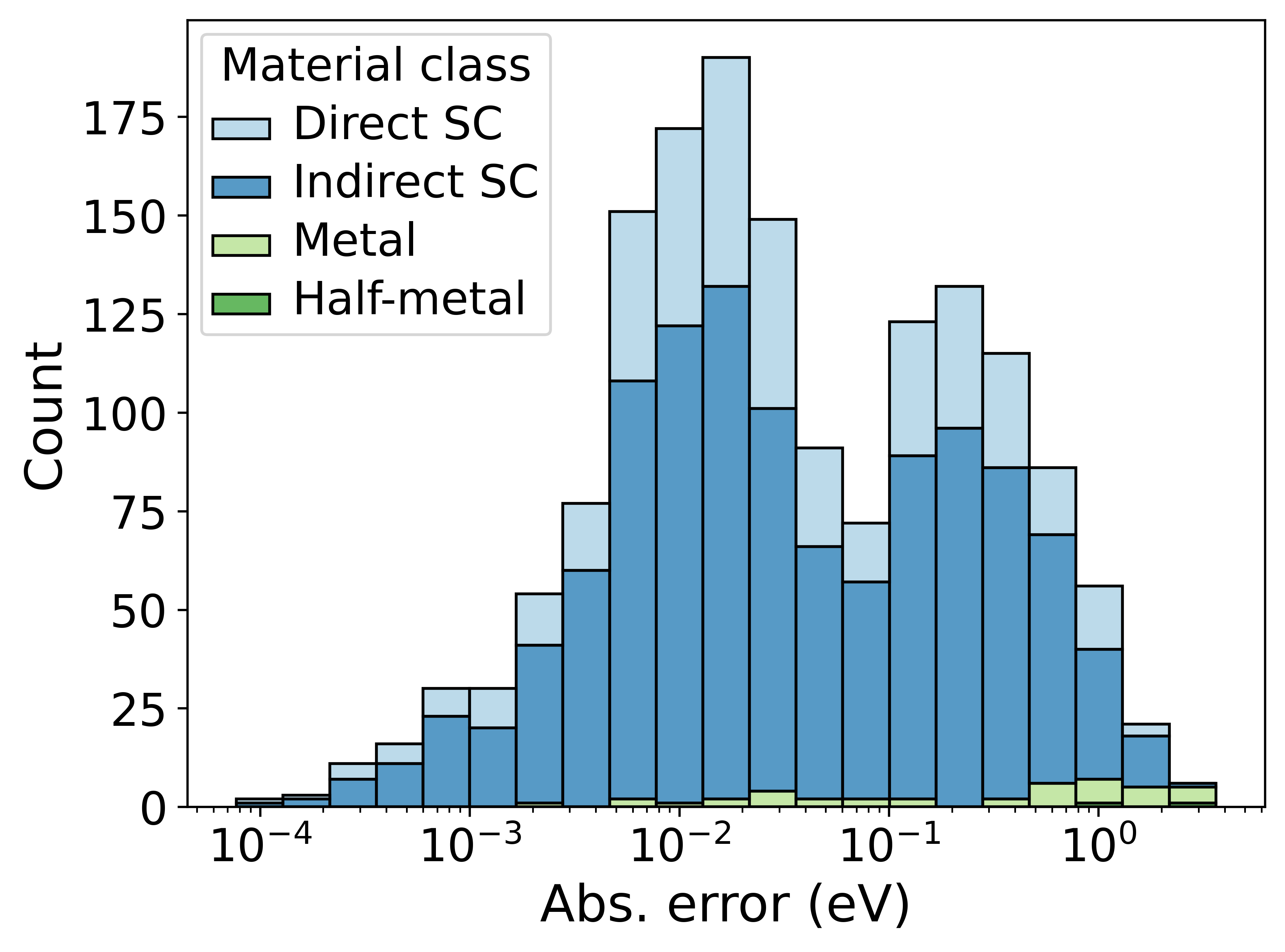}
\caption{Distribution of absolute errors for different material classes, when predicting band gaps using the ensemble NAS model. Materials are grouped into direct semiconductors (direct SC), indirect semiconductors (indirect SC), metals, and half-metals. These classes refer to the true classes of the data, not the classifier's prediction.}
\label{fig:error_hist_bandgap}
\end{figure}

Uncertainty quantification is also provided by the MPEU models. For the individual NAS models, we perform MCD while for the ensemble model we get the variance in the predictions of the models in the ensemble. For band gap regression, we find that the uncertainty of the ensemble has a correlation of 0.63 with the absolute error. In general, the uncertainty underestimates the true error, which is a known problem with uncertainly quantification in neural networks~\cite{hirschfeld2020uncertainty}. The MCD method of the best NAS model performs much worse with a correlation of 0.38. To better understand the problem of uncertainty quantification, we look in Fig. \ref{fig:error_hist_bandgap} at the distribution of absolute errors of the ensemble model. In the test data set, 60\% of the absolute errors are below 50 meV. At this level, we start to approach the numerical precision of DFT-PBE bandgaps, and much of the uncertainty we are trying to predict may be irreducible, \ie aleatoric, or just noise. Similarly, for the formation energy, most of the absolute errors are below 10 meV and thus also in the range of the numerical precision of the DFT data. It may therefore not be surprising that the formation-energy models have an even worse correlation with the uncertainty of 0.53 and 0.24 for the ensemble and best NAS model, respectively.

\begin{figure}[h]
\centering
\includegraphics[scale=0.6]{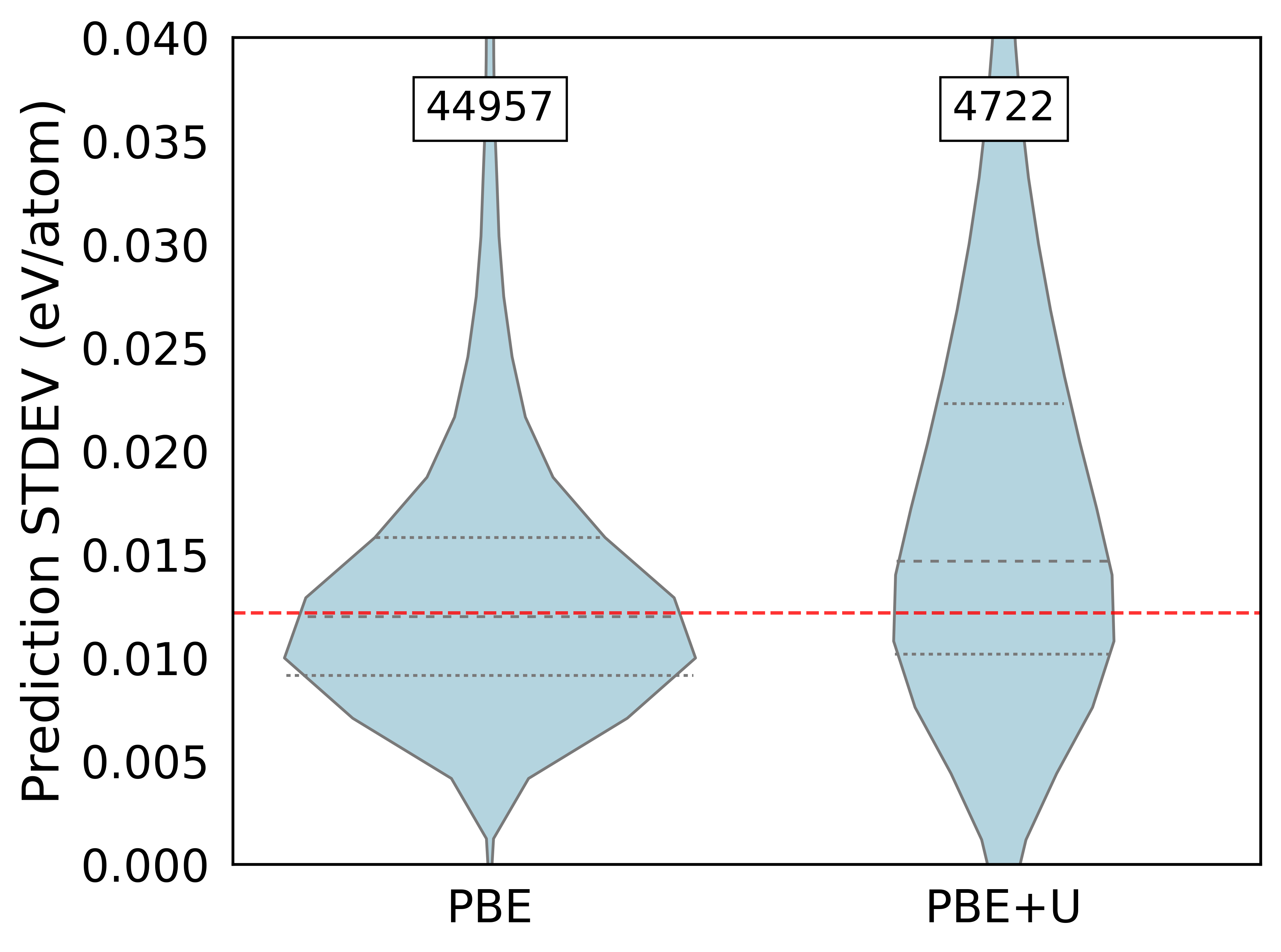}
\caption{Violin plots of the standard deviations obtained by Monte-Carlo Dropout when predicting formation energies of AFLOW data obtained by either PBE or PBE+$U$. The red horizontal line shows the median standard deviations of the predictions over the whole test split, dashed lines show quartiles. The numbers of training examples are shown at the top.}
\label{fig:mcd DFT+U formation energy}
\end{figure}

Despite the lower correlation of the MCD uncertainty estimates, they provide the user with some insight on the model's behavior for different inputs. For formation energies, the MCD uncertainty is well correlated to whether the simulation was performed with a Hubbard $U$ correction or not. This correction is applied to strongly correlated materials where the PBE functional is known to perform poorly. In the AFLOW database, PBE+$U$ is used for systems with $d$ and $f$ bands where electron localization occurs via the splitting of the energy levels of these orbitals~\cite{calderon2015aflow}. The impact of this method on the learning is seen in Fig. \ref{fig:mcd DFT+U formation energy} where the violin plots show the distribution of the model's uncertainty estimate with respect to the $U$ correction. The width of each curve corresponds with the empirical probability (\ie relative frequency) of the magnitude of the inference uncertainty. As the vast majority of test data has been  obtained without the Hubbard correction, our MPEU model appears to be more uncertain about data including it. As the $U$ parameter is an ad-hoc correction, the corresponding results may not be as systematic as the others (that still have an intrinsic error). In contrast, when the MCD method is employed for the prediction of band gaps, the median standard deviation (center of the boxplot) is lower for materials with the Hubbard-U correction (see appendix for the plot). This higher confidence of the model for band gaps performed with the Hubbard $U$ correction in comparison to the results for formation energies with the correction, indicates that the latter data are less consistent, as described in the AFLOW database~\cite{esters2023afloworg,note_AFLOW} These observations are supported by the fact that also the absolute errors depend on the regression task. Formation energies (band gaps) are worse (better) predicted by the ensemble model for materials where the correction is applied.

Finally, we want to understand why the MPEU models work so well for both the band-gap and formation-energy regressions. To do so, we remove the edge updates in our algorithm, making it equivalent to the SchNet model~\cite{schutt2018schnet}. For our implementation of SchNet we use a reduced latent size of 64, as was done in Ref.~\cite{schutt2018schnet}. This reduction in model size is also needed in our case in order to converge the validation loss during training. We observe that SchNet performs superior to ElemNet~\cite{jha2018elemnet} but falls short with respect to the MPEU models, supporting the hypothesis that edge updates increase the learning capacity of message-passing models significantly. The results are included in Table~\ref{table:1}. 

\section{Summary and conclusions}\label{sec12}

In conclusion, we find that our NAS yields ensemble models that significantly outperform models from the literature in terms of band-gap and formation-energy regression. We find that the reference model~\cite{jorgensen2018neural} applied in our context performs well for band-gap classification, being superior to the PLMF model~\cite{isayev2017universal}. The best individual NAS model does not improve over the reference model on the test split. Our analysis shows that the reference model performs significantly better on the test split as compared to the validation split, while our best NAS model yields similar results for both splits. We demonstrate the superiority of graph-based models over existing models in the literature. To improve the NAS, one could opt to use more complex search algorithms that are more exploitative (\eg genetic or Bayesian optimization). This could, however, degrade the performance of the ensemble NAS model since more exploitative search algorithms will likely return less diverse top-ranked architectures. 

The uncertainty of the models has also been analyzed. The absolute errors of our ensemble models being mostly below 50 meV and 10 meV for band-gap and formation-energy regression, respectively, approach the numerical precision of DFT results. For band-gap regression, we find a significant correlation (of 0.63) in the uncertainty for the ensemble. This also applies for data points including a Hubbard-U correction. For band gaps, the corresponding model is more certain and also less error-prone, while for formation energies, the trend is the opposite. We find that the ensemble model performs well for cubic structures but less well for triclinic materials where there are fewer training samples. We find oxide predictions to be anomalous when performing band-gap classification despite their relative abundance in the dataset. More work is required to explain this trend. Our findings may help to better understand when to apply such models and to motivate researchers to create balanced datasets with respect to structures and compositions. 

Possible future applications of our work include material discovery by exploring much larger data spaces. The NAS and ensemble methods applied to MPEU models may also be used to explore more intricate material properties such as elastic, thermal, or transport properties. Additionally, the uncertainty that the model provides, may be used in an active-learning framework. Overall, our findings may motivate other researchers to employ this methodology and our code in very different applications beyond materials science.

\backmatter

\bmhead{Code and Data Availability}

The MPEU models, code to perform the NAS, and data splits can be found online in this GitHub repository: \url{https://github.com/tisabe/jraph_mpeu}.

\bmhead{Acknowledgments}
Work carried out in part by the Max Planck Graduate Center for Quantum Materials. D.S. acknowledges support by the IMPRS for Elementary Processes in Physical Chemistry. Partial funding is appreciated from the European Union’s Horizon 2020 research and innovation program under the grant agreement Nº 951786 (NOMAD CoE) and the German Science Foundation (DFG) through the CRC FONDA, project 414984028. We are greatful to Salman Hussein for his input on the training of the ElemNet model. We thank Nakib Protik, Martin Kuban, Marcel Langer, Matthias Rupp, and Luca Ghiringhelli for fruitful discussions.

\section*{Declarations}
We declare no conflicts of interest.

\begin{appendices}
\section{}


\renewcommand{\thefigure}{A\arabic{figure}}
\renewcommand{\theHfigure}{A\arabic{figure}}
\setcounter{figure}{0}

\begin{figure}[h]
    \centering
        \centering
        \includegraphics[scale=0.30]{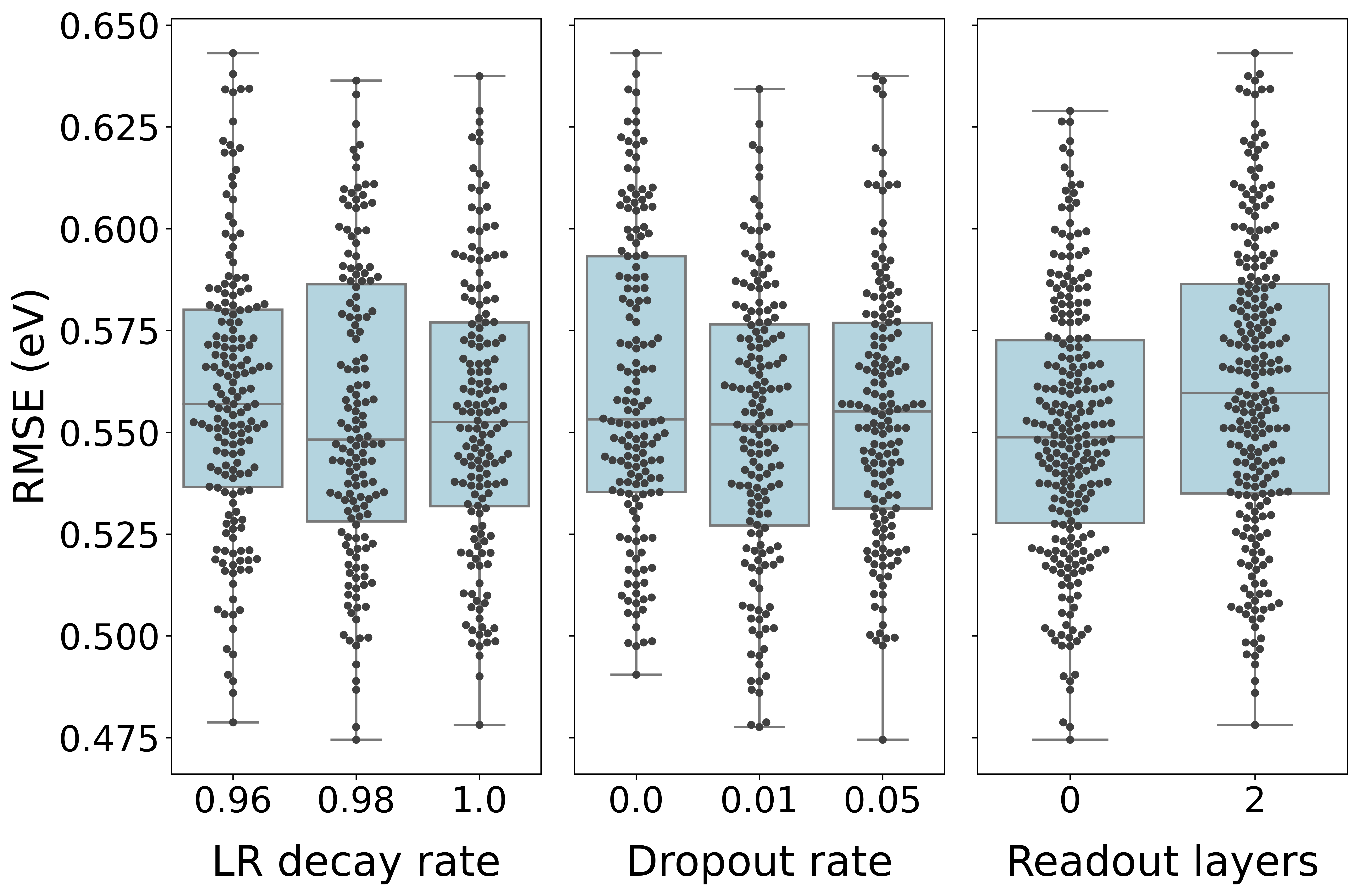}
    \vfill
        \centering
        \includegraphics[scale=0.30]{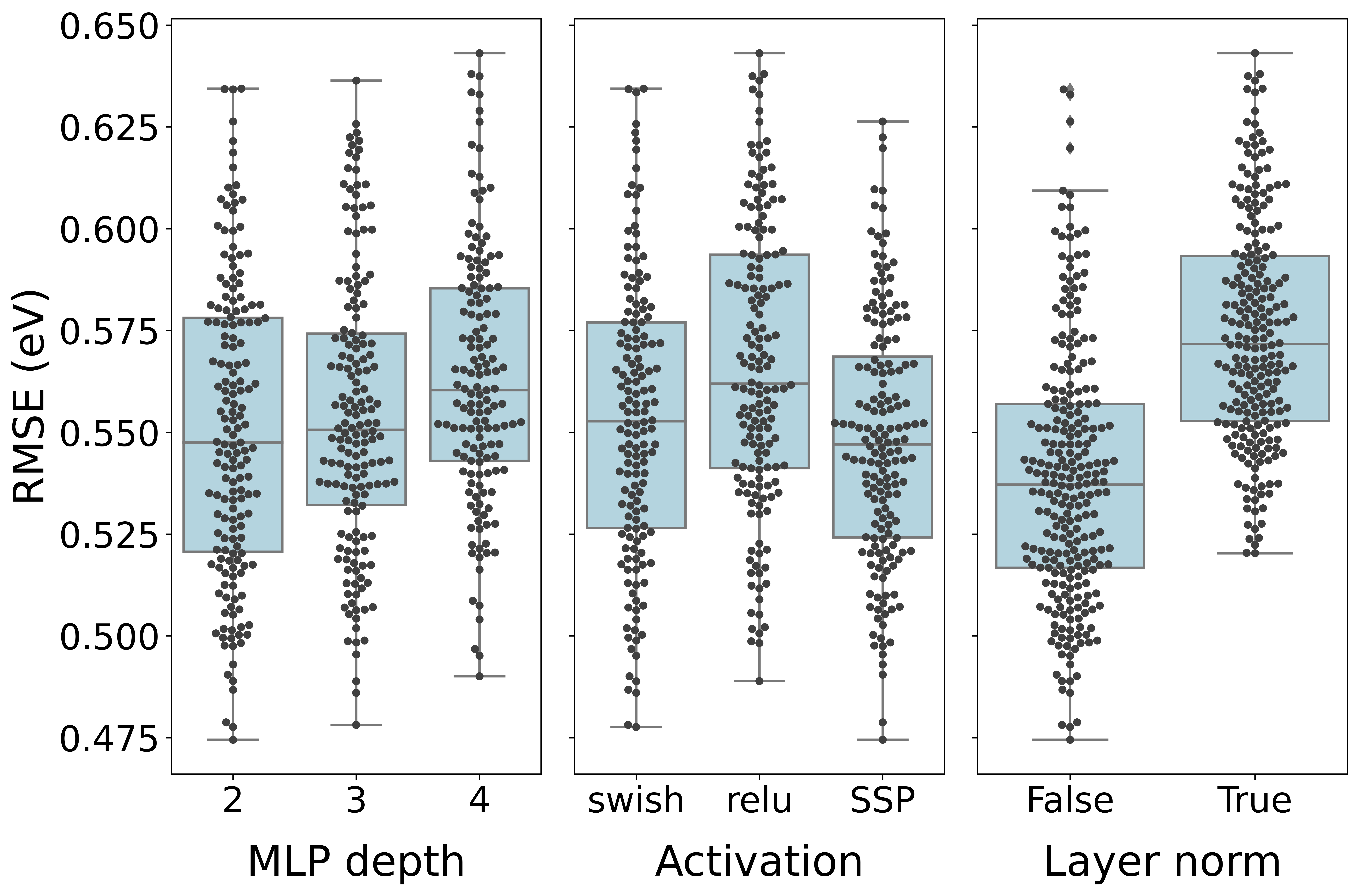}
    \caption{RMSE of the band-gap-regression task for different hyperparameters of the architecture search. The corresponding results for the most significant parameters are shown in Fig.~\ref{fig:grid search aflow bandgap}.}
    \label{fig:grid_search_aflow_1_bandgap}
\end{figure}

The plots shown here serve to better understand the model performance. We can see in Fig. \ref{fig:grid_search_aflow_1_bandgap} how different architecture parameters, not shown in Fig. \ref{fig:grid search aflow bandgap}, affect the band-gap model metrics.

\begin{figure}[h]
\centering
\centering
\includegraphics[scale=0.30]{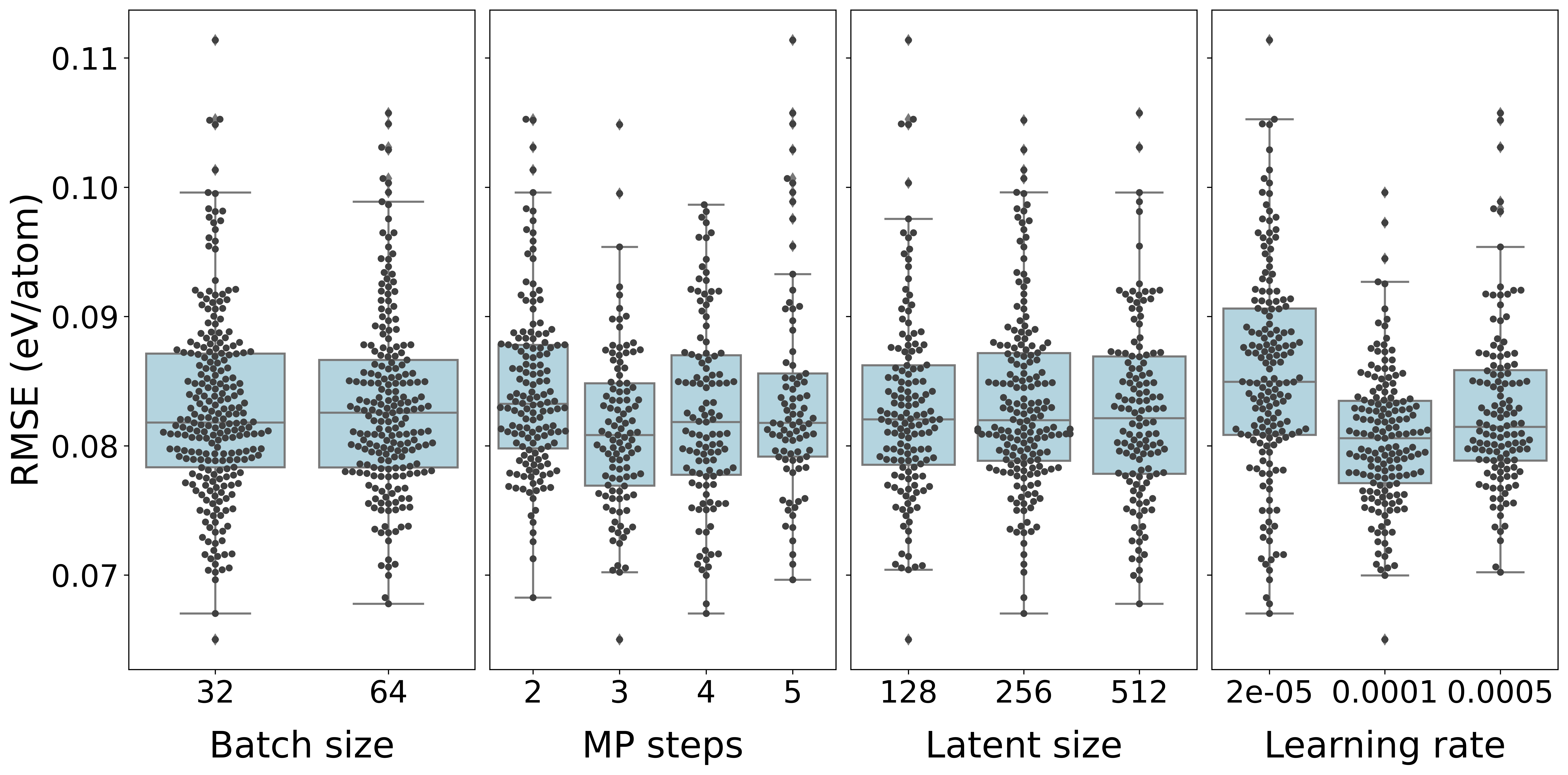}
\vfill
\centering
\centering
\includegraphics[scale=0.30]{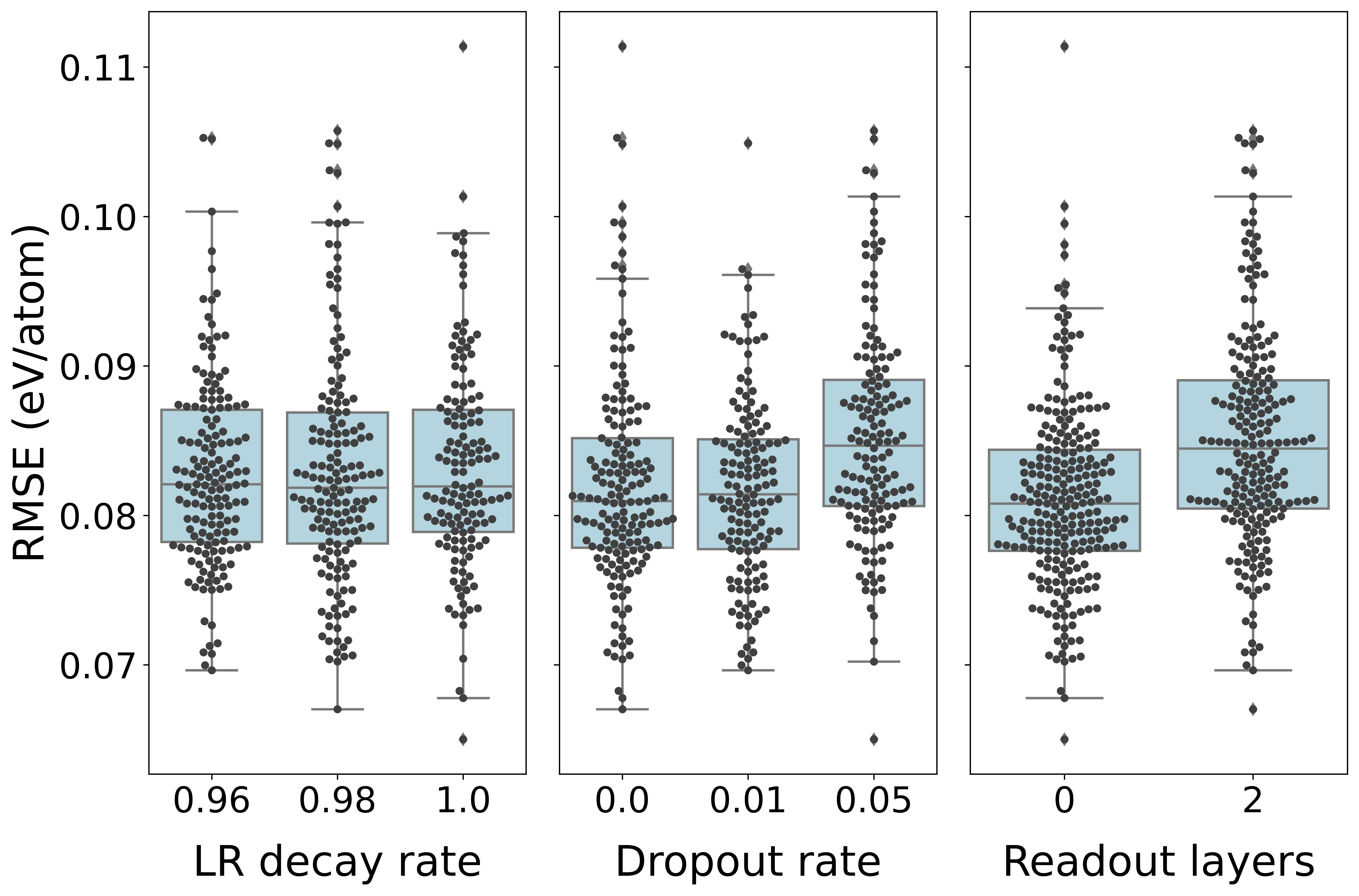}
\vfill
\centering
\includegraphics[scale=0.30]{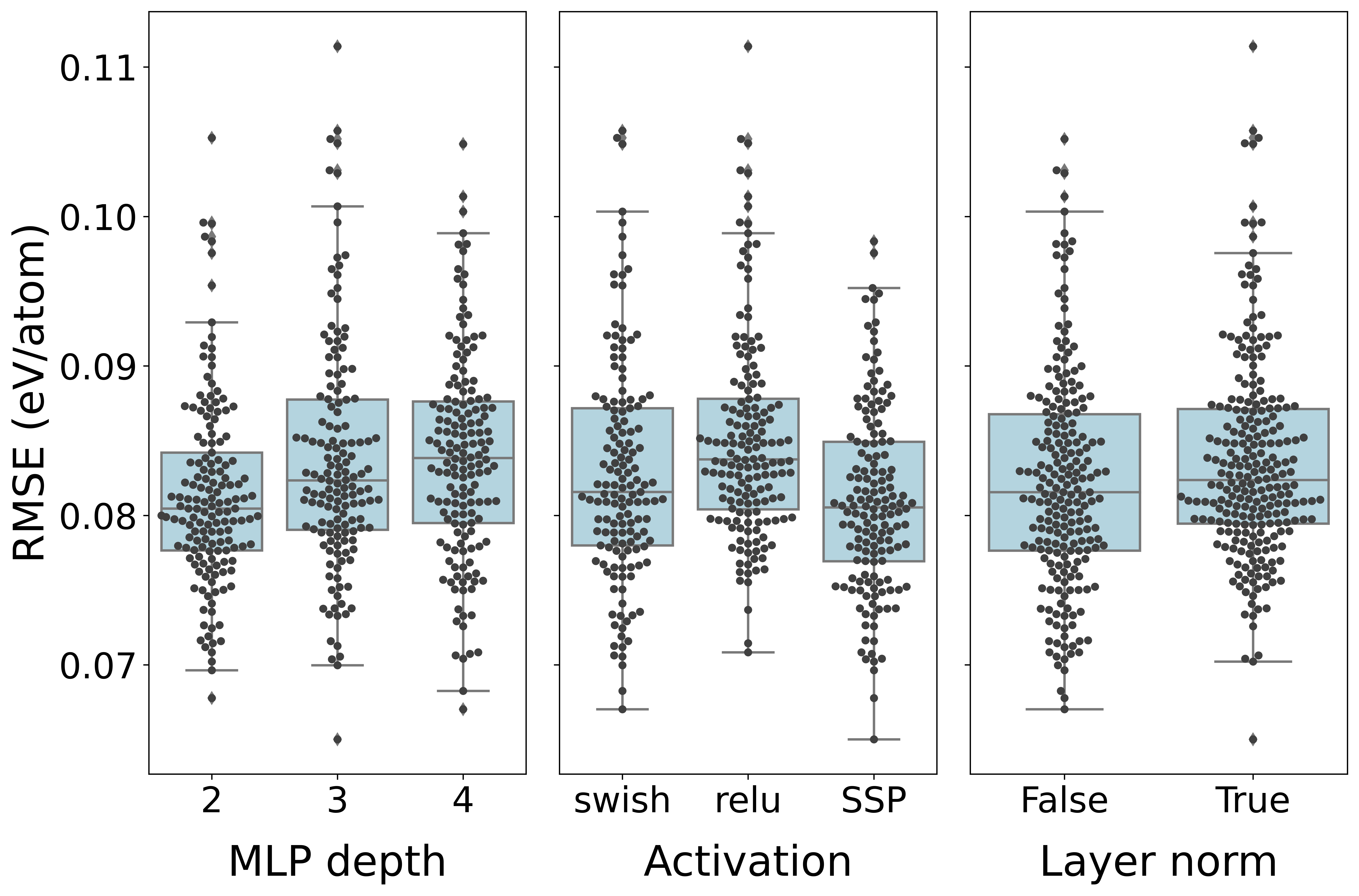}
\caption{Result of the neural-architecture search for the best model that targets formation energies (in eV/atom) of AFLOW materials. The RMSE on the validation split is shown for several parameters and settings.}
\label{fig:grid_search_aflow_1_formation_energy}
\end{figure}

The NAS results for the formation-energy task with respect to architecture and numerical parameters are depicted in Fig. \ref{fig:grid_search_aflow_1_formation_energy}.

\begin{figure}[h]
\centering
\includegraphics[scale=0.40]{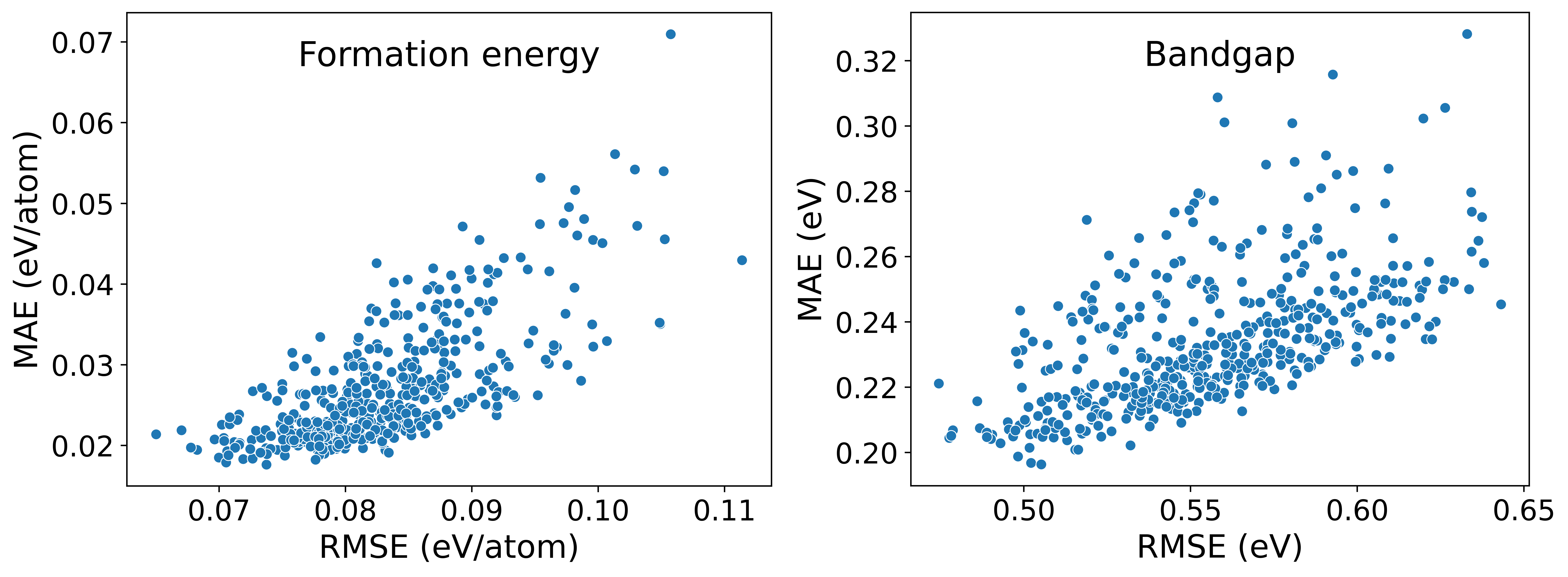}
\caption{Distribution of mean-absolute errors (MAE) as a function of the RMSE for models in the architecture search.}
\label{fig:rmse_mae_both}
\end{figure}

Fig. \ref{fig:rmse_mae_both} shows that MAE and RMSE are well correlated with each other. During our NAS training we made the assumption that we can train our models to minimize the RMSE and evaluate them on the MAE. This plot proves our assumption to be correct, \ie that the two variables are positively correlated. Minimizing the RMSE during training is easier than the MAE since the absolute value has a discontinuous derivative.

\begin{figure}[h]
\centering
\includegraphics[scale=0.6]{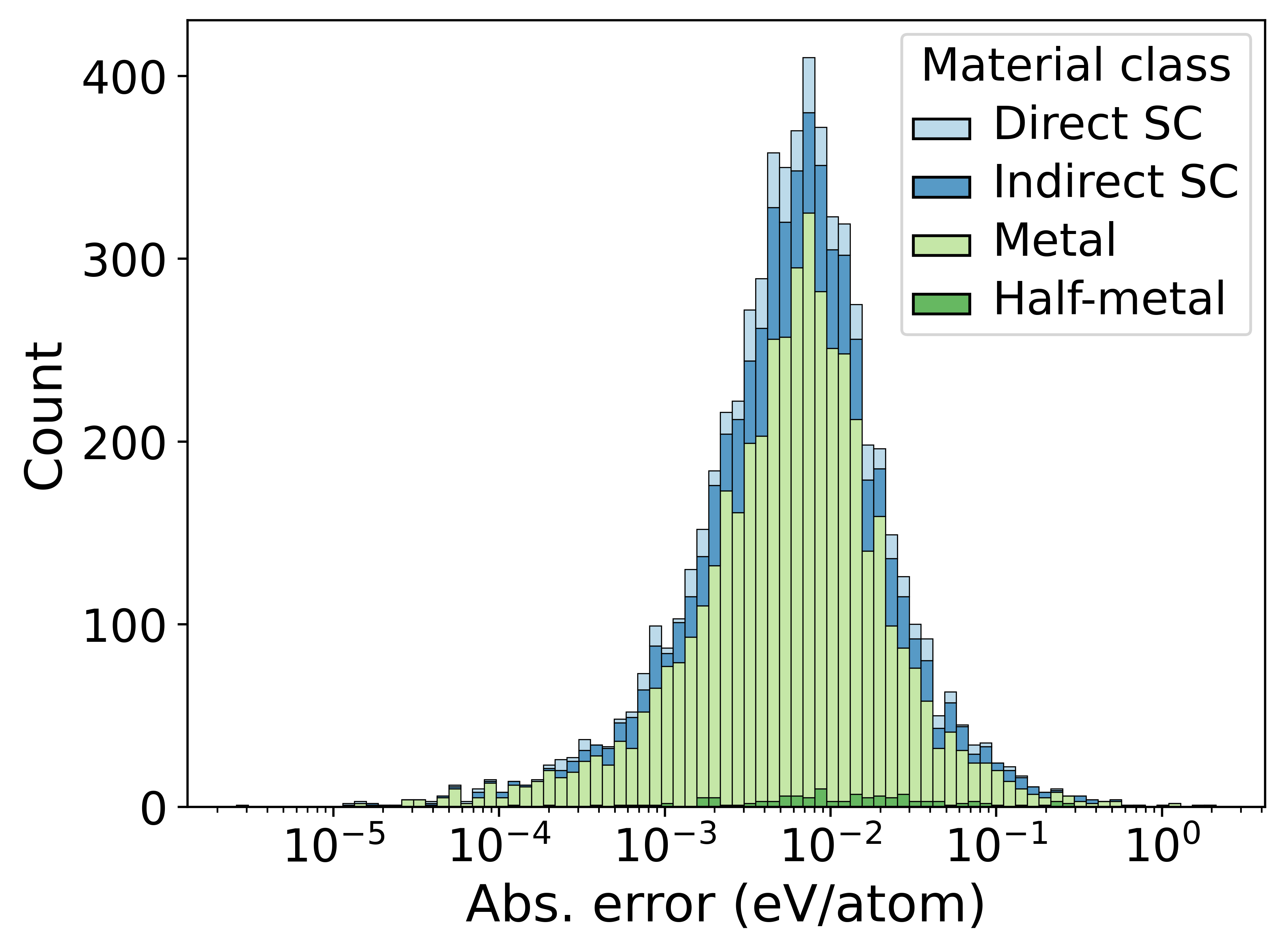}
\caption{Distribution of absolute errors for different material classes in predicting formation energies for the AFLOW dataset, using the MPEU ensemble.}
\label{fig:error_hist_formation_energy}
\end{figure}

Fig. \ref{fig:error_hist_formation_energy} shows the distribution of the absolute errors in the formation energy in the AFLOW dataset.

\begin{figure}[h]
\centering
\includegraphics[scale=0.6]{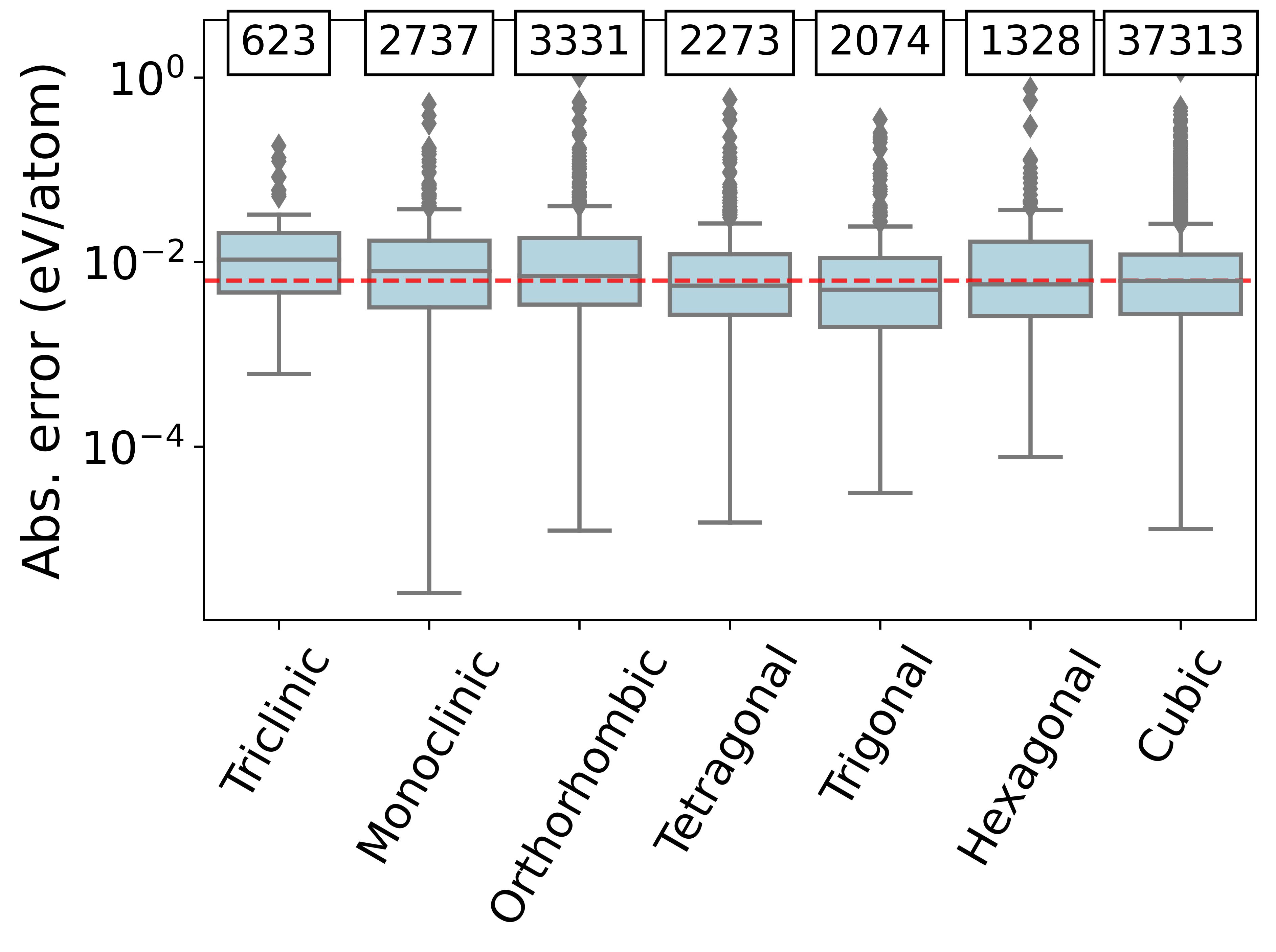}
\caption{Distribution of absolute errors for different crystal systems in predicting formation energies for the AFLOW dataset, using the MPEU ensemble. The number of training structures for each crystal system is displayed at the top. The dashed red line shows the overall median error.}
\label{fig:error_vs_crystal_formation_energy}
\end{figure}

The distribution of the absolute errors in the formation-energy models as a function of the crystal structure can be seen in Fig. \ref{fig:error_vs_crystal_formation_energy}.

\begin{figure}[h]
\centering
\includegraphics[scale=0.6]{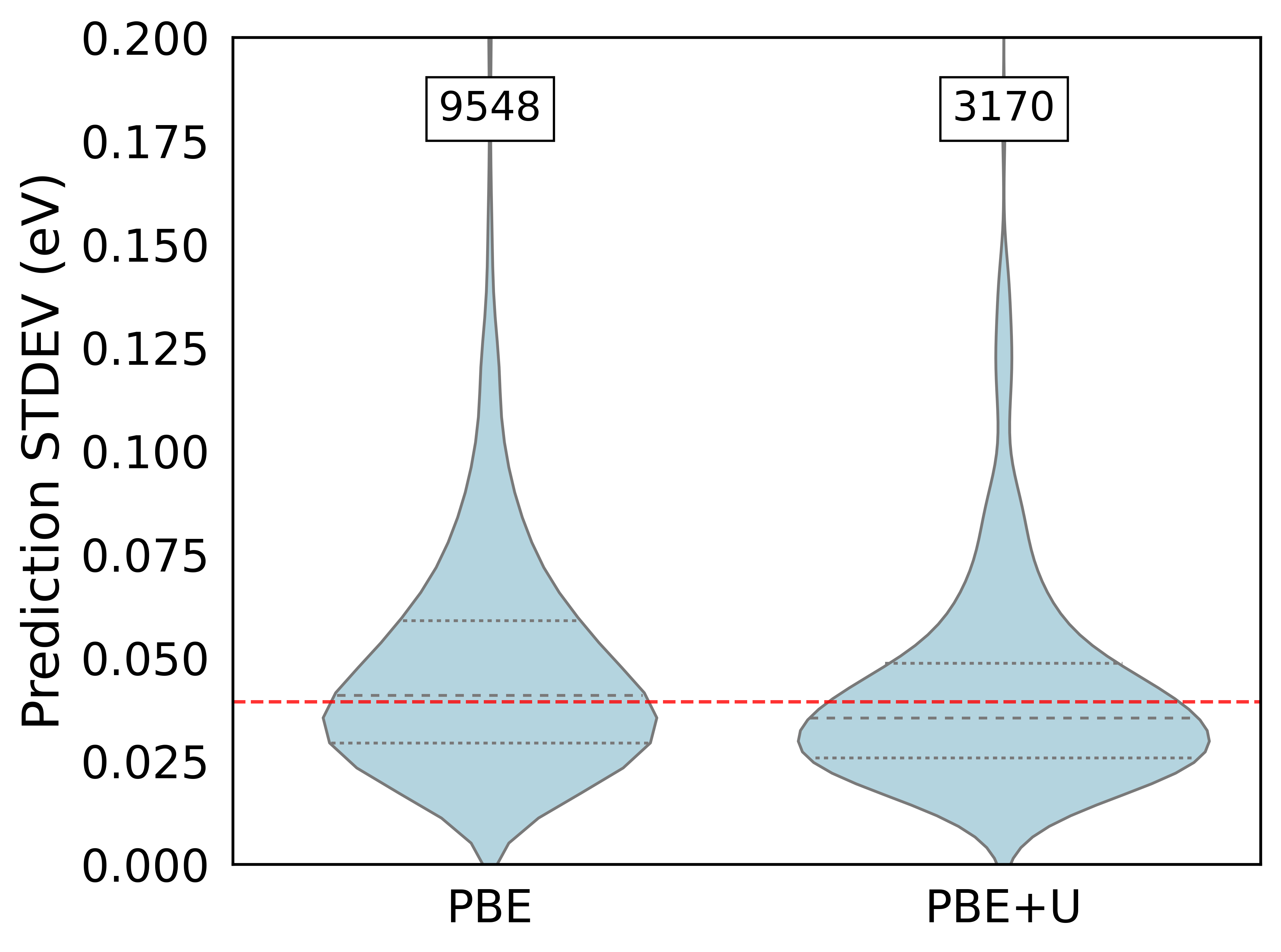}
\caption{Violin plots of the standard deviations obtained by Monte-Carlo Dropout when predicting band gaps of identified non-metals, obtained by either PBE or PBE+$U$. The red horizontal line shows the median standard deviation over the whole test split, dashed lines show quartiles. The numbers of training examples are shown at the top.}
\label{fig:mcd DFT+U bandgap}
\end{figure}

The distribution of the MC dropout uncertainty estimates are shown for the best NAS band-gap MPEU regressor in Fig. \ref{fig:mcd DFT+U bandgap}.

\begin{figure}[h]
\centering
\includegraphics[scale=0.6]{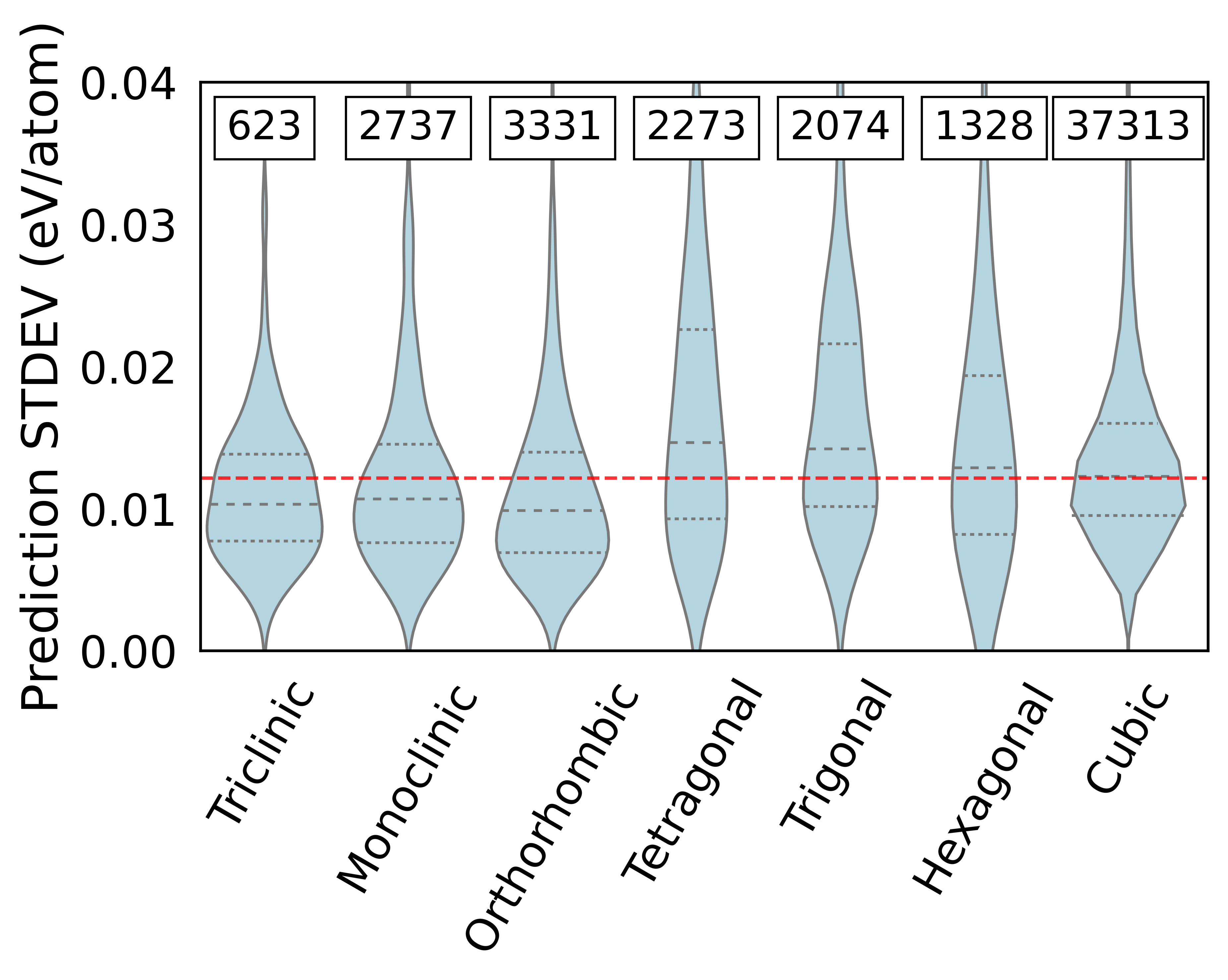}
\caption{Violin plot of the standard deviations obtained by the Monte-Carlo Dropout when making formation-energy inferences. The numbers on top indicate the numbers of materials in the training split exhibiting the respective symmetry.}
\label{fig:mcd crystal formation energy}
\end{figure}

The standard deviations from the best NAS model trained on formation energies are analyzed using Monte-Carlo Dropout. The result can be seen in Fig. \ref{fig:mcd crystal formation energy}.

\begin{figure}[h]
\centering
\includegraphics[scale=0.6]{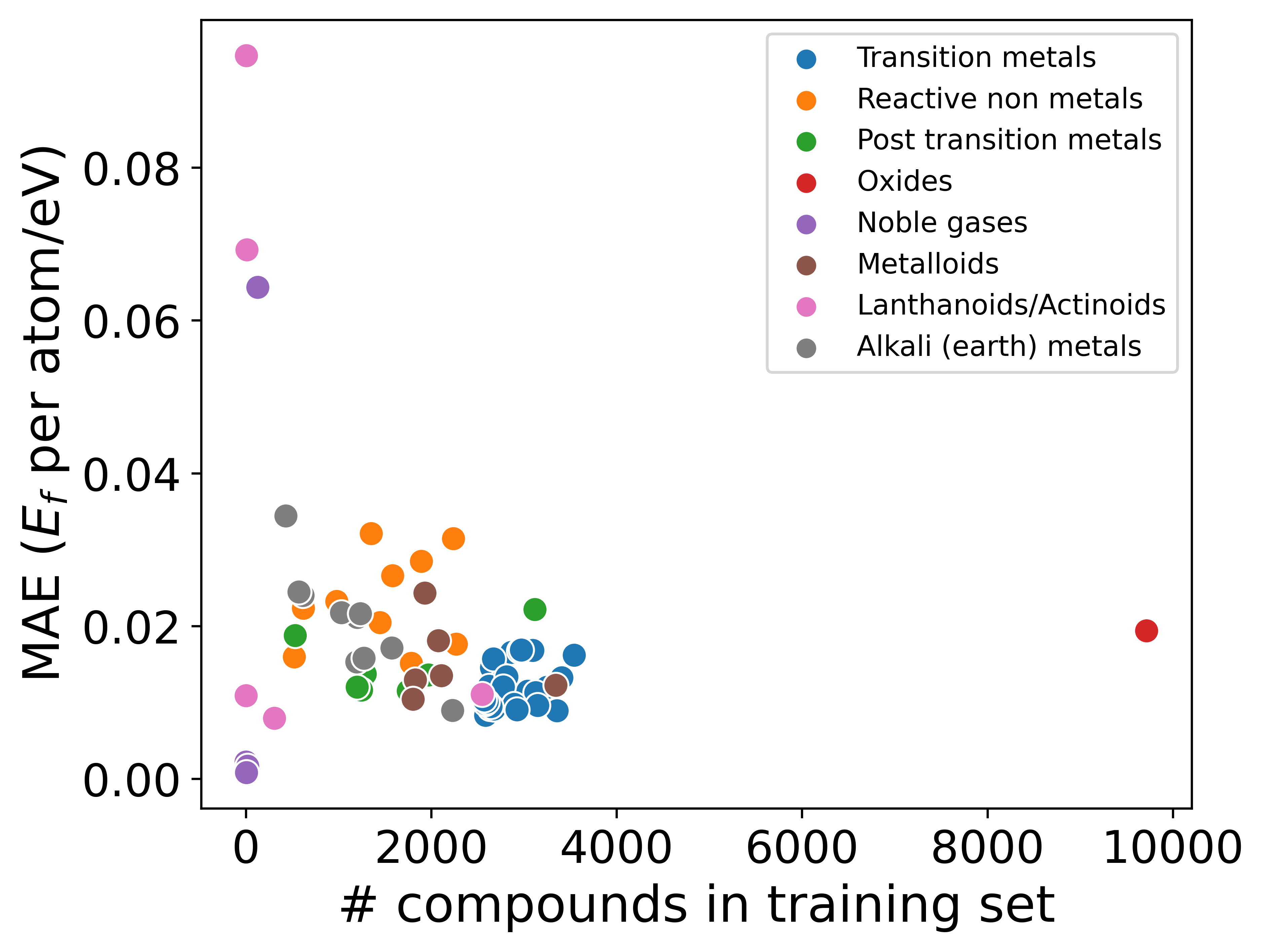}
\caption{Mean absolute errors when predicting energy of formation, depending on number of materials in each material class present in the training split.}
\label{fig:error_vs_species_ef}
\end{figure}

\begin{figure}[h]
\centering
\includegraphics[scale=0.6]{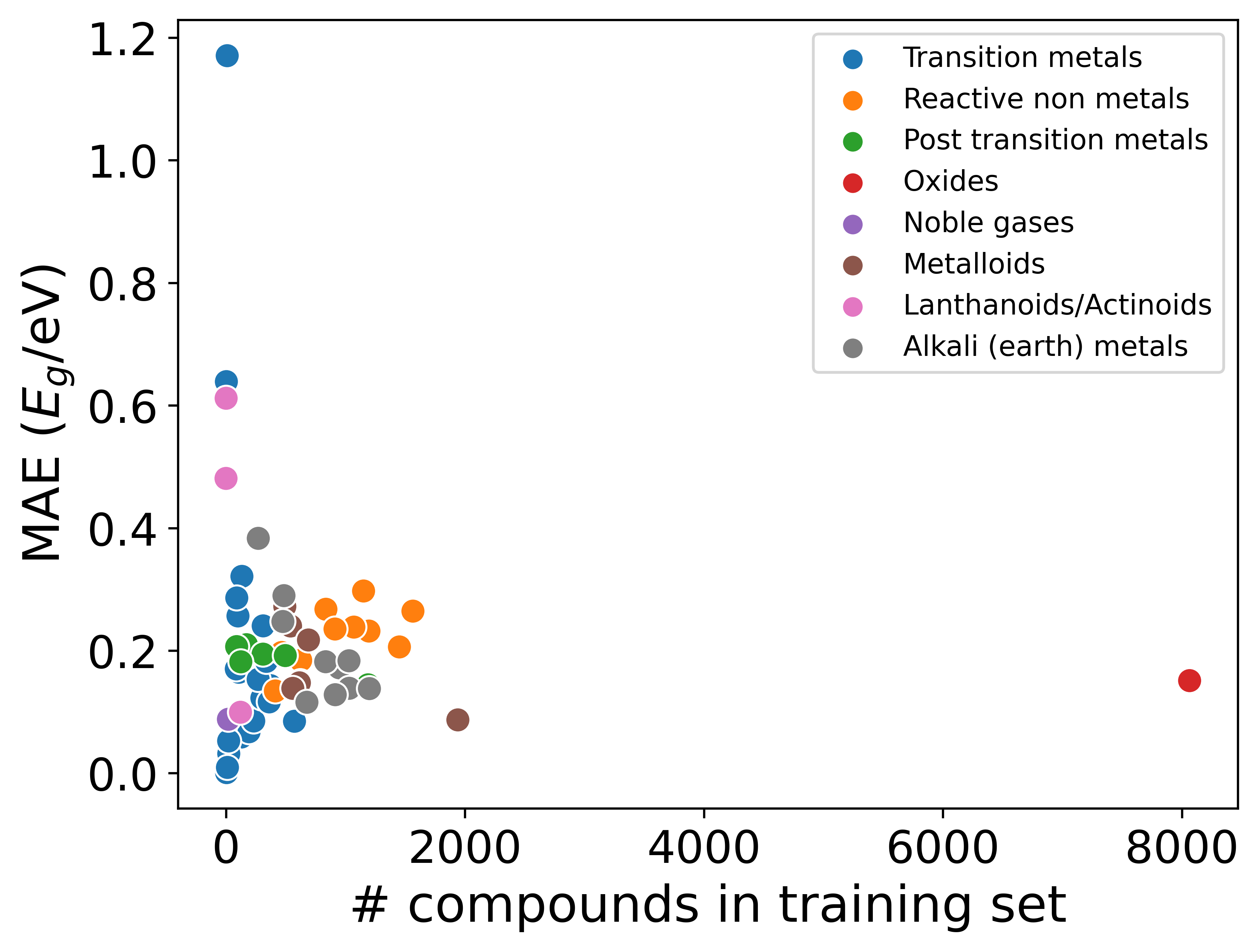}
\caption{Mean absolute errors when predicting band gaps, depending on number of materials with each species in the training split.}
\label{fig:error_vs_species_egap}
\end{figure}

The ensemble NAS model's performance on different materials as a function of the material class is shown in Fig. \ref{fig:error_vs_species_ef} for the formation-energy task and in Fig. \ref{fig:error_vs_species_egap} for band-gap regression. In both figures, we see that despite oxides being the majority class of materials in our dataset, they are not the best performing class in our dataset.

\begin{table}[h]
\centering
\caption{Regression metrics for different models, trained on AFLOW, showing their performance on the validation and test splits.}
\fontfamily{cmr}\selectfont
\begin{tabular}{p{2.2cm} p{2cm} p{1.3cm} p{1.3cm} p{1.3cm} p{1.3cm}}
\toprule
 \textbf{Property} & \textbf{Model} & \textbf{RMSE} & &
 \textbf{MAE} & \\
 &  &  Validation & Test & Validation & Test \\
 \midrule
 $E_g$ $[$meV$]$  & Ensemble & 434 & 379 & 183 & 168 \\
 & Best in NAS & 468 & 469 & 208 & 205 \\
 & Reference~\cite{jorgensen2018neural} & 506 & 399 & 209 & 180 \\
 \midrule
 $E_f$ $[$meV/atom$]$ & Ensemble & 62.5 & 56.3 & 15.7 & 15.0 \\
 & Best in NAS & 65.4 & 65.4 & 21.7 & 21.0 \\
 & Reference~\cite{jorgensen2018neural} & 75.0 & 57.5 & 19.1 & 17.9 \\
 \bottomrule
\end{tabular}
\label{table:validation}
\end{table}
\pagebreak
\end{appendices}

\bibliography{sn-bibliography}
\end{document}